\documentclass[preprint,showkeys,secnumarabic,amsfonts,showpacs,amsmath,amssymb]{revtex4}
\usepackage[dvips]{color}
\usepackage{array}
\usepackage{rotating}
\usepackage{epsfig}
\usepackage{amsmath}
\usepackage{graphicx}

\begin{document}
\title{ A new approach in stability analysis: case study: tachyon cosmology with non-minimally coupled scalar field-matter}

\author{H. Farajollahi}
\email{hosseinf@guilan.ac.ir} \affiliation{Department of Physics,
University of Guilan, Rasht, Iran}
\author{A. Salehi}
\email{a.salehi@guilan.ac.ir} \affiliation{Department of Physics,
University of Guilan, Rasht, Iran}
\date{\today}

\begin{abstract}
 \noindent \hspace{0.35cm}

 We study the general properties of attractors in a cosmological model with tachyonic potential and a scalar field non-minimally coupled to matter. In the conventional approach to the stability analysis the qualitative properties of the equations
and of the long-term behavior of the solutions are investigated where the scaling solutions are the late time
attractors and independent of the initial conditions. In this article, in a new approach, we examine
the stability analysis of the model by simultaneously solving the dynamical system and
best fitting the stability parameters with the observational data. The advantage of this
approach is that the model which was of a purely mathematical nature become physically
motivated. The number of the new critical points and also their properties, given in terms of
the best fitted parameters, may alter due to the best fitting procedure. In a further step in
stability analysis, we best-fit both the stability parameters and initial conditions with the
observational data. The results impose more constraints on the trajectories in the phase
space and provides more information about the dynamics of the cosmological model.

\end{abstract}

\pacs{04.50.Kd; 98.80.-k}

\keywords{Tachyon, stability, phase space, distance modulus, statefinder, drift velocity.}
\maketitle

\section{introduction}

Recently, the observations of high redshift type Ia supernovae and the surveys of clusters
of galaxies reveal the universe accelerating expansion and that the density of matter
is very much less than the critical density \cite{Tonry}--\cite{Spergel}. Whereas the above observational data properly complete each other,
 the dynamical dark energy (DE) proposal as an interesting possibility may arise to explain the observational constraints \cite{Sahni1}.

The two fine tuning and cosmic coincidence problems are the most serious issues with regards to the DE models and
of the most frequently used approach to moderate these problems is
the tracker field DE scenario by employing scalar field models which exhibit scaling solutions \cite{Easson}. Tracker models are independent of initial conditions used for field evolution but
require the tuning of the slope of the scalar field potential. During the scaling regime, the
scalar field energy density is of the same order of magnitude as the background energy density. The scaling solutions
as dynamical attractors can considerably resolve the two above mentioned problems. By investigating the nature of scaling solutions, one can determine whether such behavior
is stable or just a transient feature and explore the asymptotic behavior of the scalar field
potential \cite{Kim}. In studying the scalar field models with exponential potential there exist scaling attractor solutions \cite{Copeland}.

In quintessence dark energy model, there are two scaling solutions. One, the fluid-scalar field scaling solution, which remains subdominant for most of the cosmic
evolution. It is necessary that the scalar field mimics the background energy density (radiation/
matter) in order to respect the nucleosynthesis constraint and can also solve the
fine-tuning problem of initial conditions. Two, the scalar field dominated scaling solution,
which is a late time attractor and gives rise to the accelerated expansion. Since the
fluid-scalar field scaling solution is non-accelerating, we need an additional mechanism exit
from the scaling regime so as to enter the scalar field dominated scaling solution at late
times. For the discussion on the exiting mechanism one can refer to \cite{Copeland1}.

In scalar-tensor theories \cite{Sahoo}--\cite{Nojiri3}, interaction of the scalar field with matter \cite{Setare1}--\cite{Dimopoulos} and the presence of the tachyon potential in the formalism \cite{Damouri}--\cite{Padmanabhan1} separately are used to interpret the late time acceleration. For tachyon dark energy \cite{Tsujikawa,Piazza} as well as other scalar-tensor models in which a scalar field non-minimally coupled to the matter \cite{farajollahi}, the scaling solutions have been investigated separately. In this paper, we integrate both non-minimal coupling and tachyon
 models in one to investigate the scaling solutions and late time acceleration of the universe. The scalar field in our model is coupled with the matter lagrangian and also by its presence in tachyonic potential can be regarded as tachyon field. In our model, a prefect fluid with $p_{m}=\gamma\rho_{m}$ represent the matter in the universe, and the tachyonic scalar field candidates for DE.

As discussed, in stability analysis we study the qualitative properties of
the equations and of the long-term behavior of the solutions where the scaling solutions are the late time attractors and independent of the initial conditions. Here, in a new approach, we investigate the stability analysis of our model by simultaneously solving the system of equations and best fitting the stability parameters with the observational data. The advantage of this approach is that the model which was of a purely mathematical nature become physically motivated. The properties of the critical points that are now given exactly in terms of the best fitted parameters may change due to the best fitting effect. In a further step in stability analysis, we best-fit both the stability parameters and initial conditions with the observational data. The results impose more constraints on the trajectories in the phase space and provides more information about the dynamics of the universe.

The well-known geometric variables, i.e. Hubble and deceleration parameters
at the present time are used to explain the acceleration expansion of the universe. However,
considering the increased accuracy of the observational data during the last few years and
generality of the DE models, new geometrical variables are introduced to differentiate these
models and better fit the observational data. In this regard, a cosmological diagnostic pair $\{s, r\}$, called statefinder is introduced to differentiate the expansion dynamics
with higher derivatives of the scale factor and is a natural next step beyond the well known geometric variables \cite{Sahni}. The
statefinder pair has been used to explore a series of dark energy and cosmological models,
including $\Lambda$ cold dark matter (LCDM), quintessence, coupled quintessence, Chaplygin gas, holographic dark energy
models, braneworld models, and so on \cite{Alam,Zimdahl,Yi}.

The "Cosmological
Redshift Drift" (CRD) test which maps the expansion of the universe
directly is also examined in here. We assume that the universe is
homogeneous and isotropic at the cosmological scales \cite{Lis}. The test is based on very simple and straightforward
physics. Observationally, it is a very challenging task and
requires technological breakthroughs, for more details see \cite{Cristiani}--\cite{Loeb}.

The manuscript is organized as followers. In section 2, we derive the field equations for the model. In section 3, we perform stability analysis and obtain the autonomous equations in term of the new dynamical variables for the model. The critical points are also obtained. section 3.1 is designed to solve once again the autonomous equations by best fitting the stability parameters and initial conditions. Then we obtain the new properties for the critical points. The Stability of the best fitted critical points and phase space is studied in section 3.2. In section 4.1, for the best fitted model, the cosmological parameters are investigated. In addition, in section 4.2, the model is tested against observational data for the drift velocity. In section 5 we present summary and remarks.

\section{The Model}

The model is presented by the action,
\begin{eqnarray}\label{action}
S=\int[\frac{R}{16\pi G}-V(\phi)\sqrt{1-\phi_{,\mu}\phi^{,\mu}}+f(\phi)\mathcal{L}_{m}]\sqrt{-g}dx^{4},
\end{eqnarray}
where $R$ is Ricci scalar, $G$ is the newtonian constant gravity, and the second term in the action is tachyon potential.
 The $f(\phi)$ is
an analytic function of the scalar field. The last term in the lagrangian brings about the nonminimal
interaction between the matter and the scalar field.
The variation of action (\ref{action})  with respect to the metric tensor components in a spatially flat FRW  cosmology
yields the field equations:
\begin{eqnarray}
&&3H^{2}=\rho_{m}f+\frac{V(\phi)}{\sqrt{1-\dot{\phi}^{2}}},\label{fried1}\\
&&2\dot{H}+3H^2=-\gamma\rho_{m}f+V(\phi)\sqrt{1-\dot{\phi}^{2}},\label{fried2}
\end{eqnarray}
where we put  $8\pi G=c=\hbar=1$ and $ H=\frac{\dot{a}}{a}$  with $a$ is the scale factor of the universe. we also assume a perfect fluid with $p_{m}=\gamma\rho_{m}$. Note that in here $\gamma$ is the EoS parameter for the matter field in the universe. Variation of the action (\ref{action}) with respect to the scalar field  $\phi$ provides the wave
equation for the scalar field as
\begin{eqnarray}\label{phiequation}
\ddot{\phi}+(1-\dot{\phi}^{2})(3H\dot{\phi}+\frac{V^{'}}{V})=-\frac{\epsilon f^{'}}{V}(1-\dot{\phi}^{2})^{\frac{3}{2}}\rho_{m},
\end{eqnarray}
where prime indicates differentiation with respect to $\phi$ and $\epsilon=1-3\gamma$.
From equations (\ref{fried1})--(\ref{fried2}) one arrives at the conservation equation,
\begin{eqnarray}\label{conserv2}
\dot{(\rho_{m}f)}+3H(1+\gamma)\rho_{m}f=\epsilon \rho_{m}\dot{f}.
\end{eqnarray}
From equations (\ref{fried1}) and (\ref{fried2}), by defining the effective energy density and pressure,  $\rho_{eff}$ and $p_{eff}$, one can identify an effective EoS parameter as
\begin{eqnarray}\label{omegaef}
\omega_{eff}\equiv\frac{p_{eff}}{\rho_{eff}}=\frac{\gamma\rho_{m}f-V(\phi)\sqrt{1-\dot{\phi}^{2}}}{\rho_{m}f+\frac{V(\phi)}{\sqrt{1-\dot{\phi}^{2}}}}
\end{eqnarray}
In the next section we study the stability analysis of the model in the phase space.

\section{perturbation and Stability Analysis}

The structure of the dynamical system can be studied via  phase plane analysis,
by introducing the following dimensionless variables,
\begin{eqnarray}\label{defin}
 x={\frac{\rho_{m}f}{3 H^{2}}},\ \ y=\frac{V}{3H^{2}},\ \ z=\dot{\phi},
\end{eqnarray}
and parameters
\begin{eqnarray}\label{defin2}
 \alpha={\frac{\dot{f}}{fH}},\ \ \beta=\frac{\dot{V}}{VH}.
\end{eqnarray}
 Then using (\ref{fried1})-(\ref{conserv2}), the equations for the new dynamical variables are,
\begin{eqnarray}
x'&=&-2x(-\frac{3}{2}-\frac{3\gamma}{2}x+\frac{3y^2}{2-2x})+\epsilon\alpha x-3(1+\gamma)x, \label{x1} \\
y'&=&-2y(-\frac{3}{2}-\frac{3\gamma}{2}x+\frac{3y^2}{2-2x})+\beta y \label{y1}\\
z'&=&-(1-z^{2})(3z+\frac{\beta}{z})-\epsilon(1-z^{2})^{\frac{3}{2}}\frac{xy}{z}\label{z1}
\end{eqnarray}
where prime " $'$ "in here and from now on means derivative with respect to $N = ln (a)$. By using the Fridmann constraint equation (\ref{fried1}) in terms of the new dynamical variables, i.e.
\begin{eqnarray}\label{constraint}
x+\frac{y}{\sqrt{1-z^{2}}}=1,
\end{eqnarray}
the equations (\ref{x1})-(\ref{z1}) reduce to,
\begin{eqnarray}
x'&=&-2x(-\frac{3}{2}-\frac{3\gamma}{2}x+\frac{3y^2}{2-2x})+\epsilon\alpha x-3(1+\gamma)x,\label{x11}\\
y'&=&-2y(-\frac{3}{2}-\frac{3\gamma}{2}x+\frac{3y^2}{2-2x})+\beta y.\label{y11}
\end{eqnarray}
In term of the new dynamical variable we also have,
\begin{eqnarray}
\frac{\dot{H}}{H^{2}}=-\frac{3}{2}(1+\gamma x+\frac{y^2}{x-1}).\label{hh2}
\end{eqnarray}
In stability formalism, by simultaneously solving $x'=0$, $y'=0$
 the fixed points (critical points) can be obtained.
The critical points that depend on the cosmological and stability parameters $\gamma$, $\alpha$ and $\beta$ are
illustrated in Table I.\\

\begin{table}[ht]
\caption{The fixed points} 
\centering 
\begin{tabular}{c c c c  } 
\hline\hline 
points  &  FP1  & FP2 \ & FP3   \\ [4ex] 
\hline 
$x$ & 0 & 0 & 0    \\ 
\hline 
$y$ & 0 &$ \sqrt{\frac{\beta+3}{3}}$&- $ \sqrt{\frac{\beta+3}{3}}$    \\
\hline 
\end{tabular}
\label{table:1} 
\end{table}

Substituting
linear perturbations $x'\rightarrow x'+\delta x'$, $y'\rightarrow y'+\delta y'$, about the critical points into
the two independent equations (\ref{x11})--(\ref{y11}), to the first
orders in the perturbations, gives us two eigenvalues $\lambda_{i} (i=1,2)$ which has to be negative as a requirement by stability method. In the following, the nature of the three critical points are given with the stability conditions:\\

$FP1:\lambda_{1P1}=3+\beta,\lambda_{2P1}=-3\gamma+(1-3\gamma)\alpha $\\
$FP2,3:\lambda_{1P2,3}=-6-2\beta,\lambda_{2P2,3}=-3-3\gamma+(1-3\gamma)\alpha-\beta $\\

$where$

$FP1:stable for \left\{
\begin{array}{ll}
\alpha<\frac{3\gamma}{1-3\gamma}, \beta<-3\ \ \ \gamma<\frac{1}{3}\\ \alpha>\frac{3\gamma}{1-3\gamma}, \beta<-3\ \ \ \gamma>\frac{1}{3}\\ \hbox{$ \beta<-3\ \ \ \ \ \ \ \  \ \ \ \ \ \ \ \ \gamma=\frac{1}{3}$}
\end{array}
\right.
$\\

$FP2,3:stable for \left\{
   \begin{array}{ll}
     [[\frac{3\gamma}{3\gamma-1}\leq\alpha,\beta>-3\alpha\gamma-3\gamma+\alpha-3],[\alpha<\frac{3\gamma}{3\gamma-1}, \beta>-3]]\ \ \gamma<\frac{1}{3}\\  \ [[\frac{3\gamma}{3\gamma-1}\geq\alpha,\beta>-3\alpha\gamma-3\gamma+\alpha-3],[\alpha>\frac{3\gamma}{3\gamma-1}, \beta>-3]]\ \ \gamma>\frac{1}{3}\\ \hbox{$ \beta>-3\ \ \ \ \ \ \ \  \ \ \ \ \ \ \ \ \gamma=\frac{1}{3}$}
   \end{array}
 \right.
$\\

As can be seen, all the critical points are stable for the given conditions on  $\alpha$, $\beta$ and $\gamma$.

Ont the other hand, two of the cosmological parameters which relates the dynamics of the universe with the observational data are the EoS and deceleration parameters. In terms of the new dynamical variables in our model they are given by,
\begin{eqnarray}
\omega_{eff}&=&\gamma x+\frac{y^2}{x-1},\label{eff}\\
q&=&\frac{1}{2}(1+\gamma x+\frac{y^2}{x-1}).\label{qeff}
\end{eqnarray}
Moreover, the statefinder parameters $\{r,s\}$ in terms of the dynamical variables are,
\begin{eqnarray}
r&=&\frac{\ddot{H}}{H^{3}}-3q-2=\frac{d}{dN}(\frac{\dot{H}}{H^{2}})+2(\frac{\dot{H}}{H^{2}})^{2}+3\frac{\dot{H}}{H^{2}}+1,\label{r}\\
s&=&\frac{r-1}{3(q-\frac{1}{2})},\label{s}
\end{eqnarray}
where $\frac{\dot{H}}{H^{2}}$ in terms of the new dynamical variables is given by (\ref{hh2}).
For our model these parameters are presented in Table II:\\
\begin{table}[hb]
\caption{Properties of the fixed points } 
\centering 
\begin{tabular}{c c c c c c  } 
\hline\hline 
points   & q & $\omega_{eff}$ & r & s & acceleration \\ [3ex] 
\hline 
FP1  & 1/2 & 0 & $1$&$1$& No\\ 
FP2  & $-1-\frac{\beta}{2}$ & $-1-\frac{\beta}{3}$ & $\frac{\beta^{2}}{2}+\frac{3\beta}{2}+1$&$\frac{-\beta}{3}$& $\beta>-2$ \\
FP3 & $-1-\frac{\beta}{2}$ & $-1-\frac{\beta}{3}$ &$\frac{\beta^{2}}{2}+\frac{3\beta}{2}+1$&$\frac{-\beta}{3}$& $\beta>-2 $\\
 [1ex] 
\hline 
\end{tabular}
\label{table:2} 
\end{table}\

In the following, we solve the above equations for $\gamma=0$ and  $\gamma=1/3$ by best fitting the model parameters and initial conditions with the observational data using the $\chi^2$ method. This enables us to find the solutions for the above equations and conditions for the stability of the critical points that are physically more meaningful and observationally more favored. In the next section we best fit the model with the observational data for distance modulus.

\subsection{Best fitting the stability parameters and initial conditions}

The difference between the absolute and
apparent luminosity of a distance object is given by, $\mu(z) = 25 + 5\log_{10}d_L(z)$ where the Luminosity distance quantity, $d_L(z)$ is given by
\begin{equation}\label{dl}
d_{L}(z)=(1+z)\int_0^z{\frac{dz'}{H(z')}}.
 \end{equation}
 In our model, from numerical computation one can obtain $H(z)$ which can be used to evaluate $\mu(z)$. To best fit the model for the parameters $\alpha$ and $\beta$ and the initial conditions $x(0)$, $y(0)$, $H(0)$ with the most recent observational data, the Type Ia supernovea (SNe Ia), we employe the $\chi^2$ method. We constrain the parameters including the initial conditions by minimizing the $\chi^2$ function given as
\begin{equation}\label{chi2}
    \chi^2_{SNe}(\alpha, \beta, H(0), x(0), y(0))=\sum_{i=1}^{557}\frac{[\mu_i^{the}(z_i|\alpha, \beta, H(0), x(0), y(0)) - \mu_i^{obs}]^2}{\sigma_i^2},
\end{equation}
where the sum is over the SNe Ia sample. In relation (\ref{chi2}), $\mu_i^{the}$ and $\mu_i^{obs}$ are the distance modulus parameters obtained from our model and from observation, respectively, and $\sigma$ is the estimated error of the $\mu_i^{obs}$. From numerical computation, Table III shows the best fitted model parameters for $\gamma=0$ and $\gamma=1/3$.\\

\begin{table}[ht]
\caption{Best-fitted model parameters and initial conditions.} 
\centering 
\begin{tabular}{c c c c c c c } 
\hline 
parameters  &  $\alpha$  &  $\beta$ \ & $x(0)$\ & $y(0)$\ & $H(0)$ \ & $\chi^2_{min}$\\ [2ex] 
\hline 
$\gamma=0$&$-0.17$  & $-1.62$ \ & $-0.2$\ & $-0.8$\ & $0.904$ \ & $548.8365694$ \\
$\gamma=1/3$  & $-$  & $-1.1$ \ & $-0.2$\ & $-0.8$\ & $0.909$ \ & $552.6788148$ \\
\hline 
\end{tabular}
\label{table:1} 
\end{table}\

Fig. 1) shows the constraints on the parameters $\alpha$, $\beta$ and the initial condition $H(0)$ at the $68.3\%$, $95.4\%$ and $99.7\%$ confidence levels in both cases of $\gamma=0, 1/3$. Note that in case of $\gamma=1/3$, since equation (\ref{y11}) is independent of $\alpha$, there is no constraint on $\alpha$ in this case, and therefore the equation becomes independent of this parameter as can be seen in the table III and Fig. 1.\\

 \begin{tabular*}{2.5 cm}{cc}
\includegraphics[scale=.3]{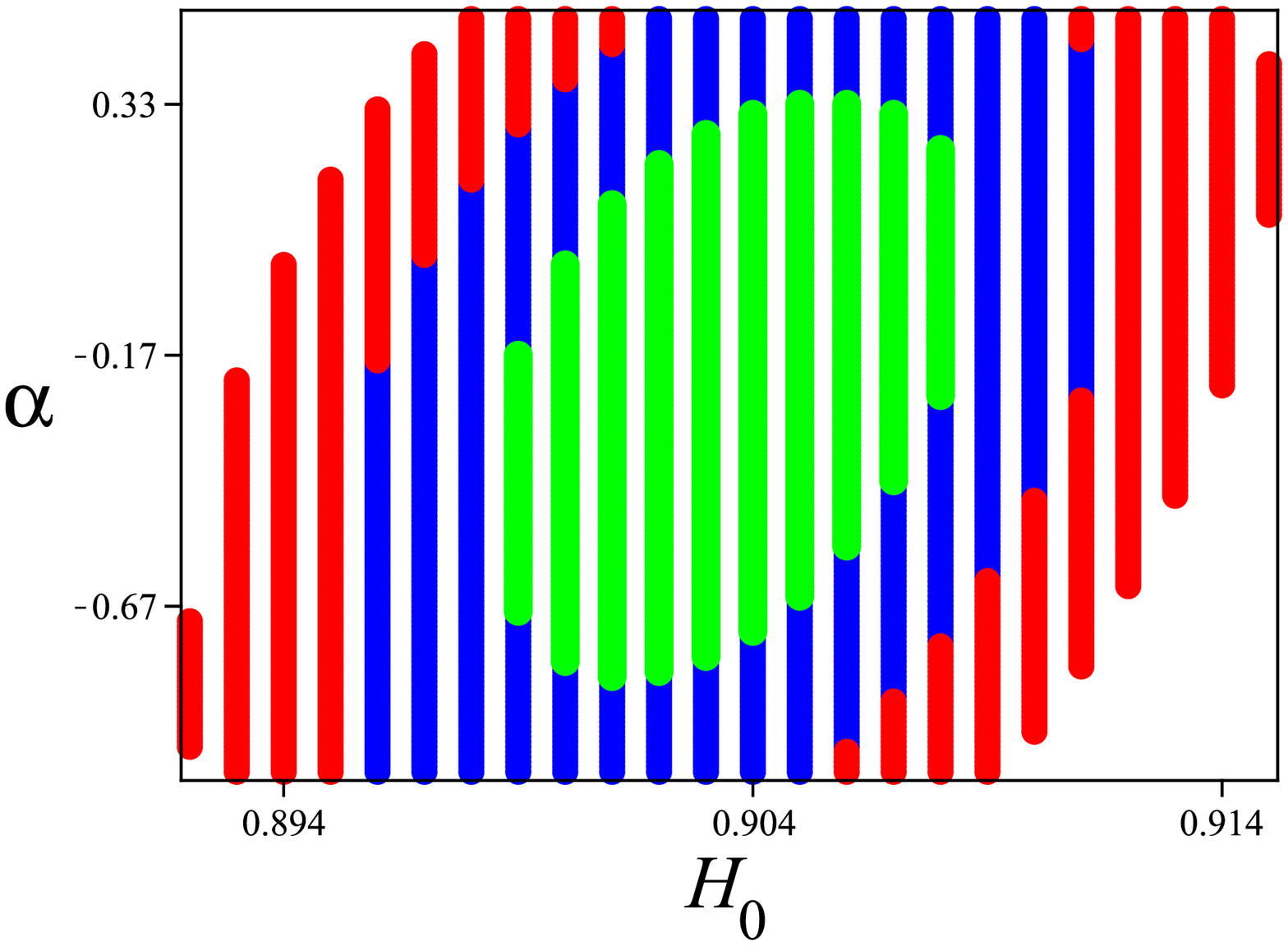}\hspace{0.1 cm}\includegraphics[scale=.3]{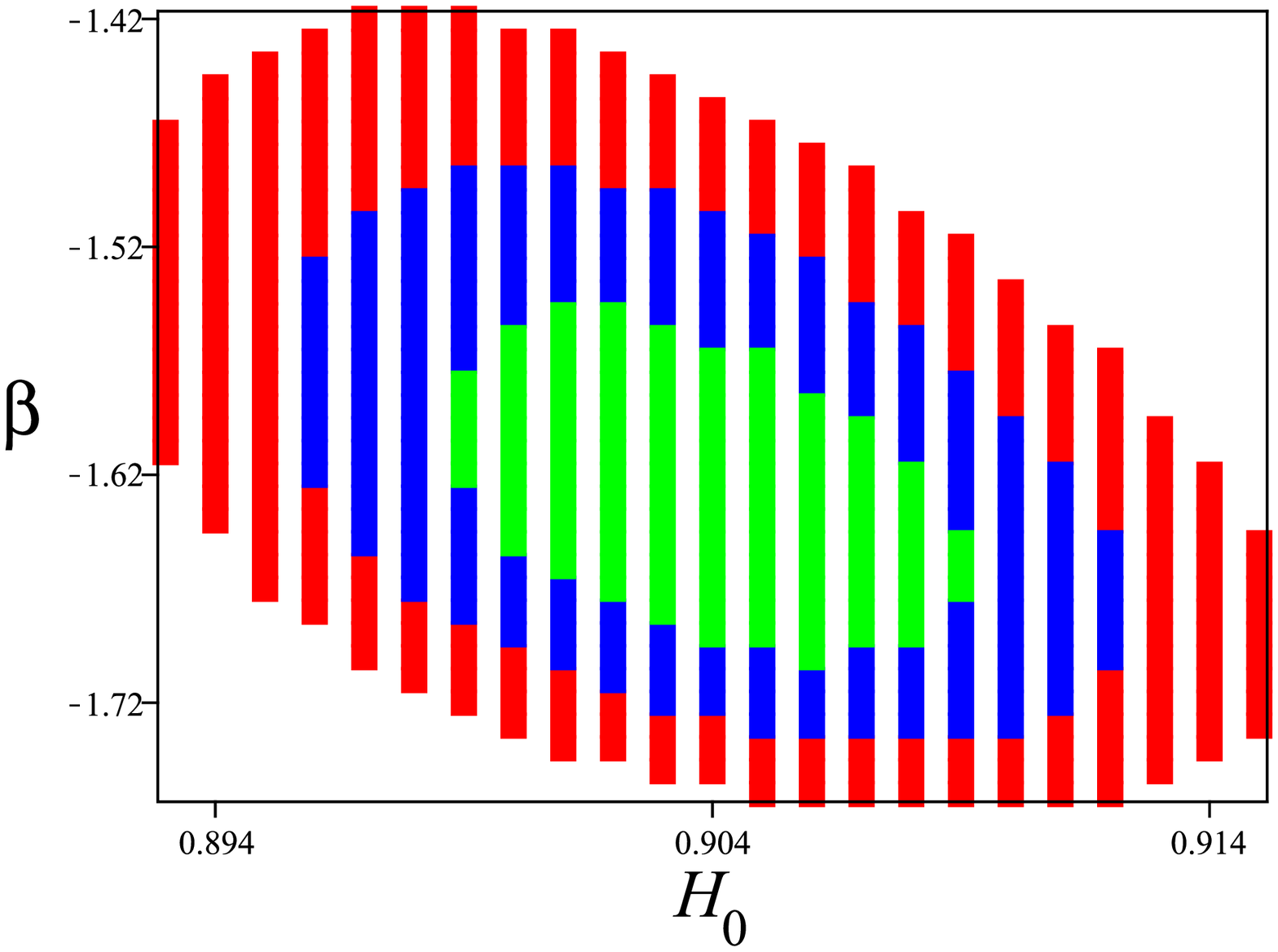}\hspace{0.1 cm}\includegraphics[scale=.3]{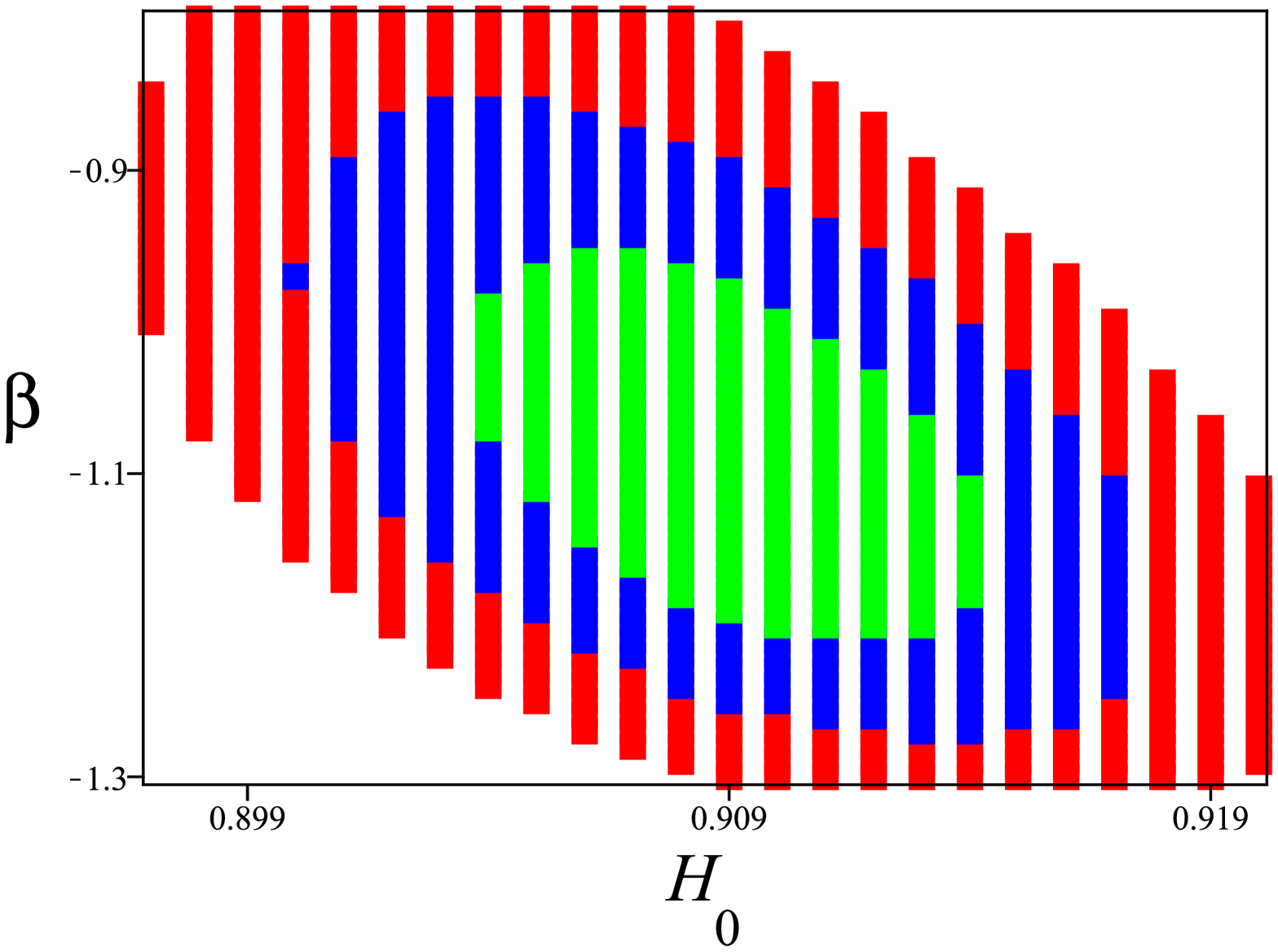}\hspace{0.1 cm}\\
Fig. 1:  The graph of confidence level for $(\alpha,H_{0})$ and $(\beta,H_{0})$ \\Left and Middle)for case $\gamma=0$, Right) for $\gamma=\frac{1}{3}$
\end{tabular*}\\

Alternatively, we can plot the likelihood for the model parameters in both cases $\gamma=0$ and $\gamma=1/3$ ( Figs. 2-4).

\begin{tabular*}{2.5 cm}{cc}
\includegraphics[scale=.28]{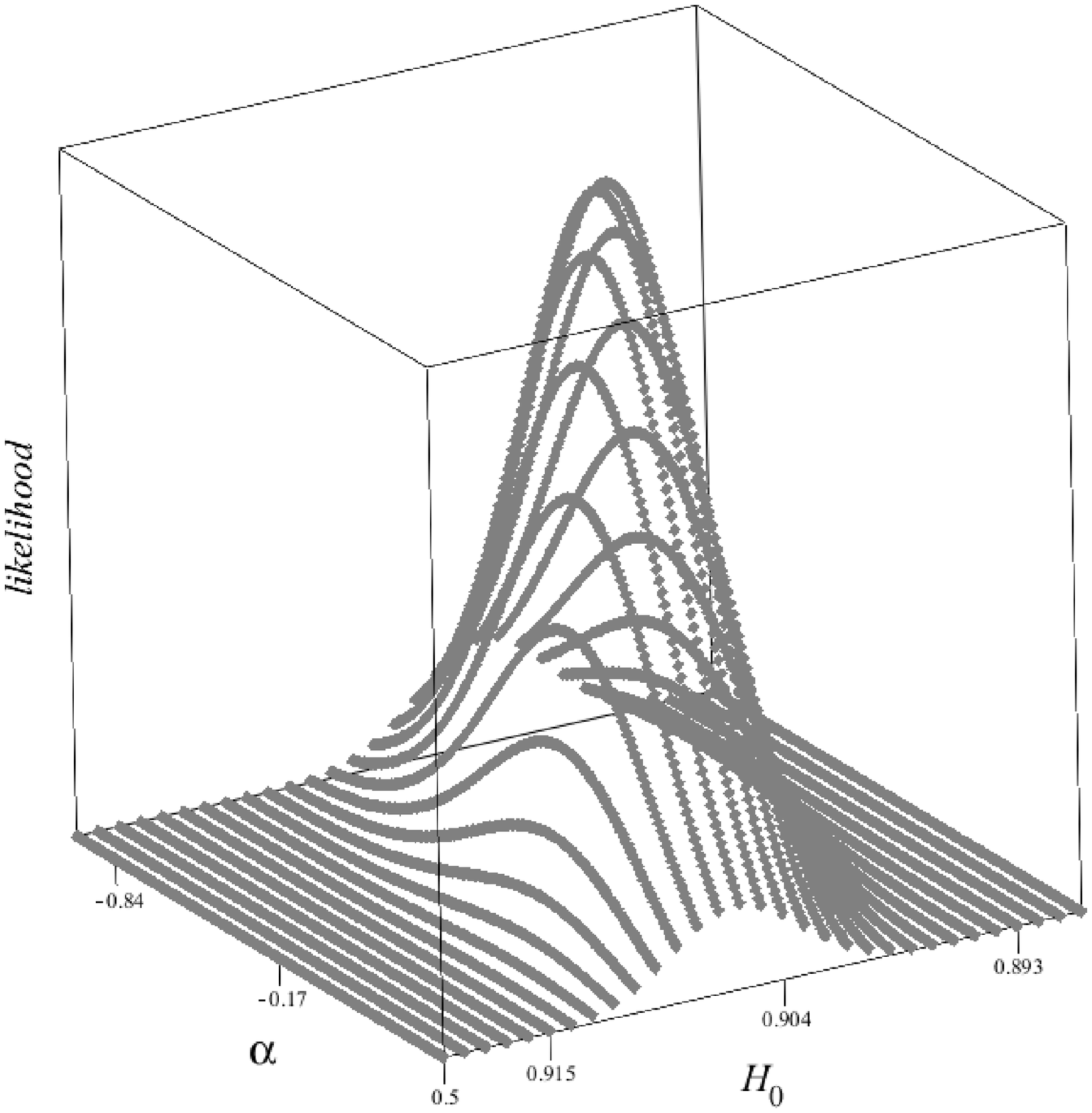}\hspace{0.1 cm}\includegraphics[scale=.3]{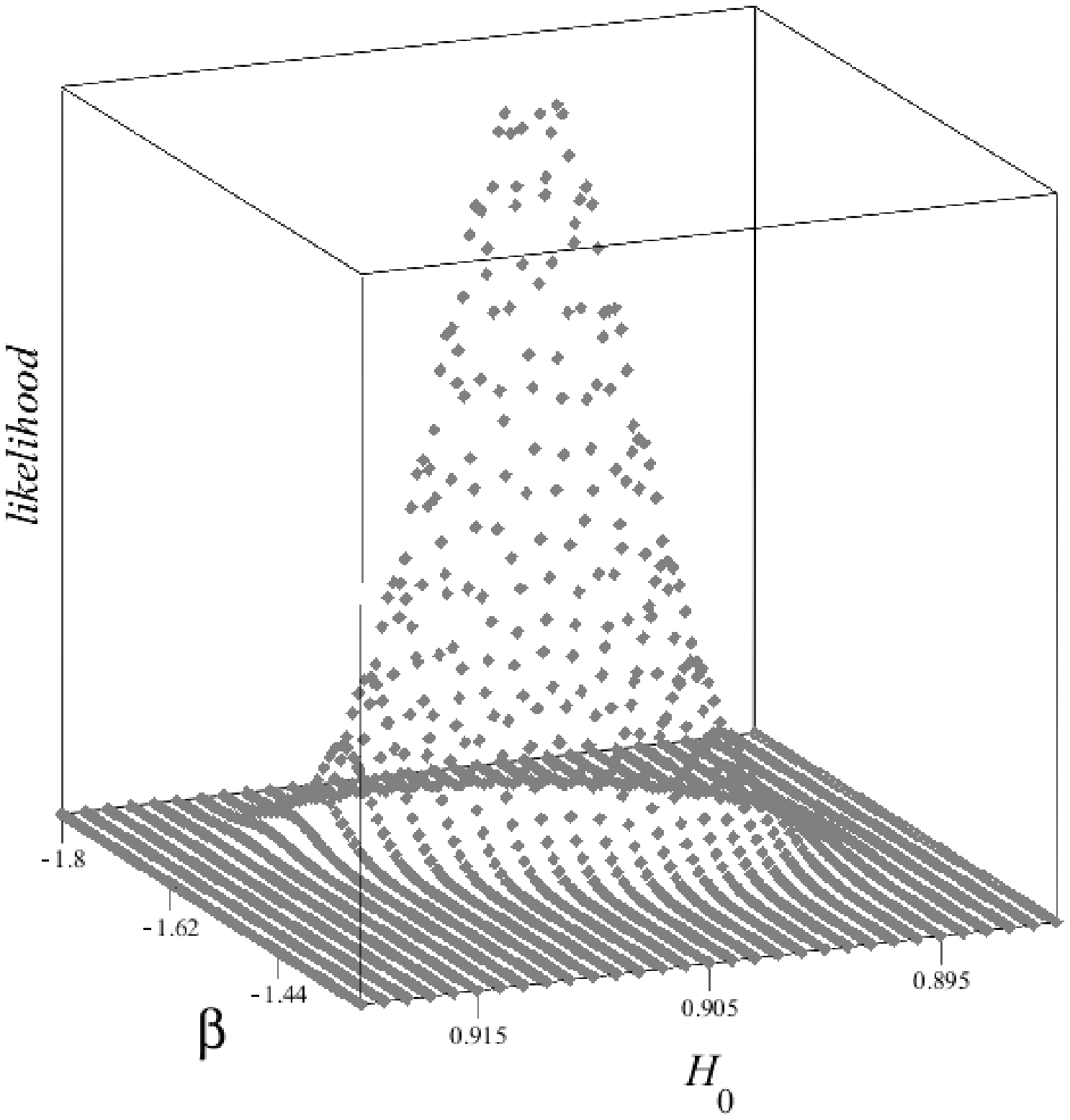}\hspace{0.1 cm}\includegraphics[scale=.3]{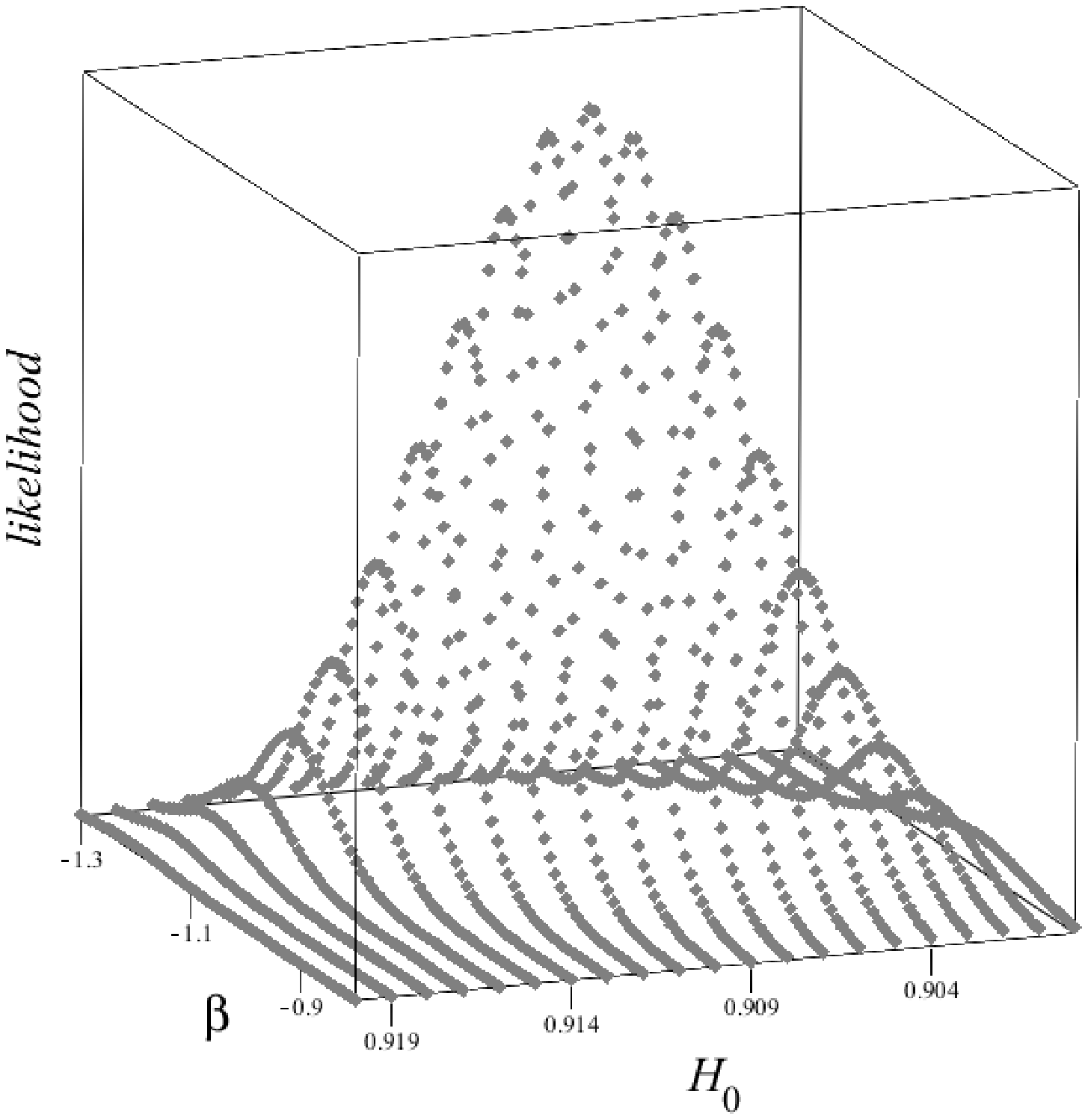}\hspace{0.1 cm}\\
Fig. 2: The graph of 2-dim likelihood distribution for $(\alpha,H_{0})$ and $(\beta,H_{0})$ \\Left and Middle)for case $\gamma=0$, Right) for $\gamma=\frac{1}{3}$
\end{tabular*}\\

\begin{tabular*}{2.5 cm}{cc}
\includegraphics[scale=.3]{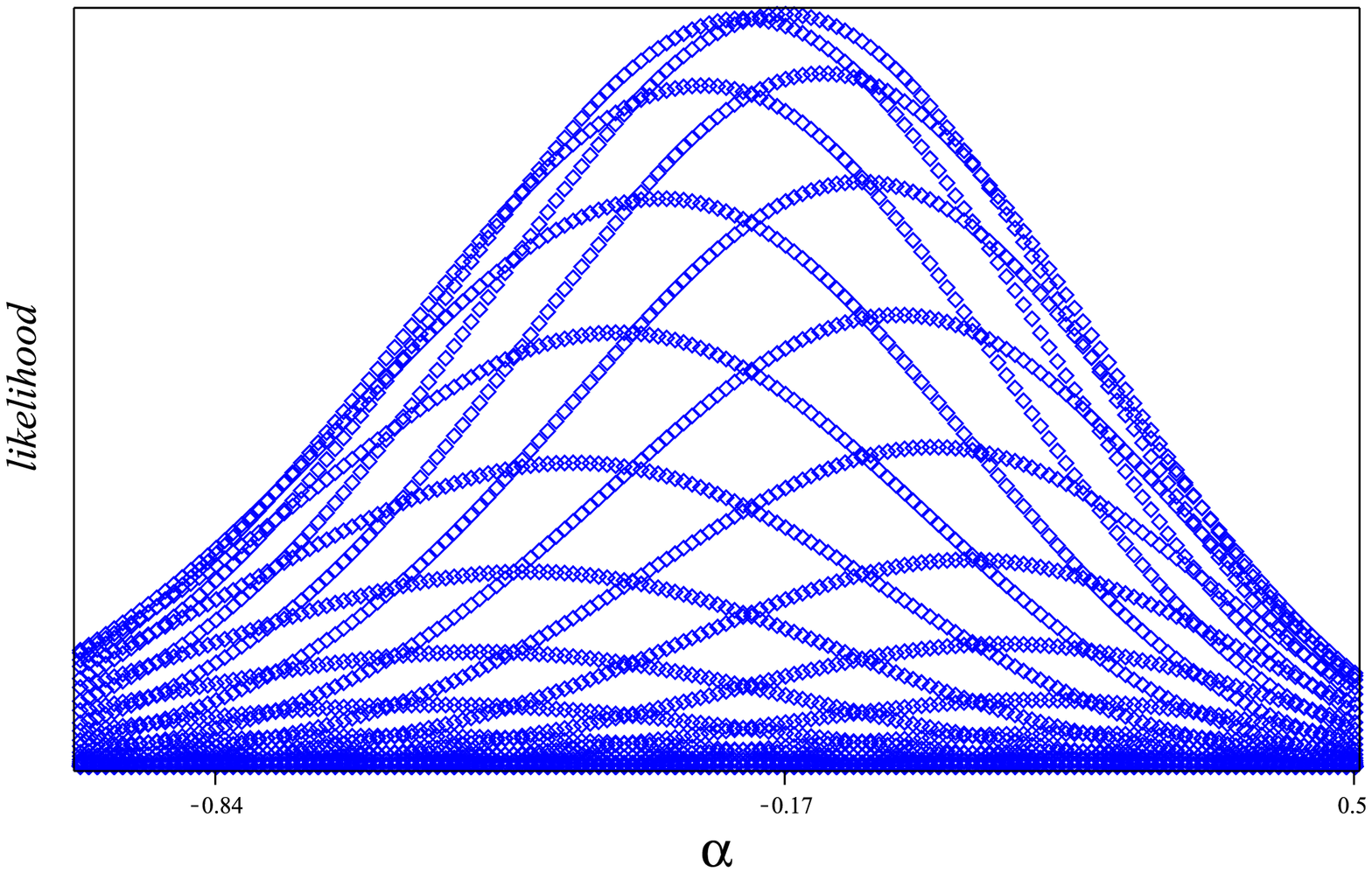}\hspace{0.1 cm}\includegraphics[scale=.3]{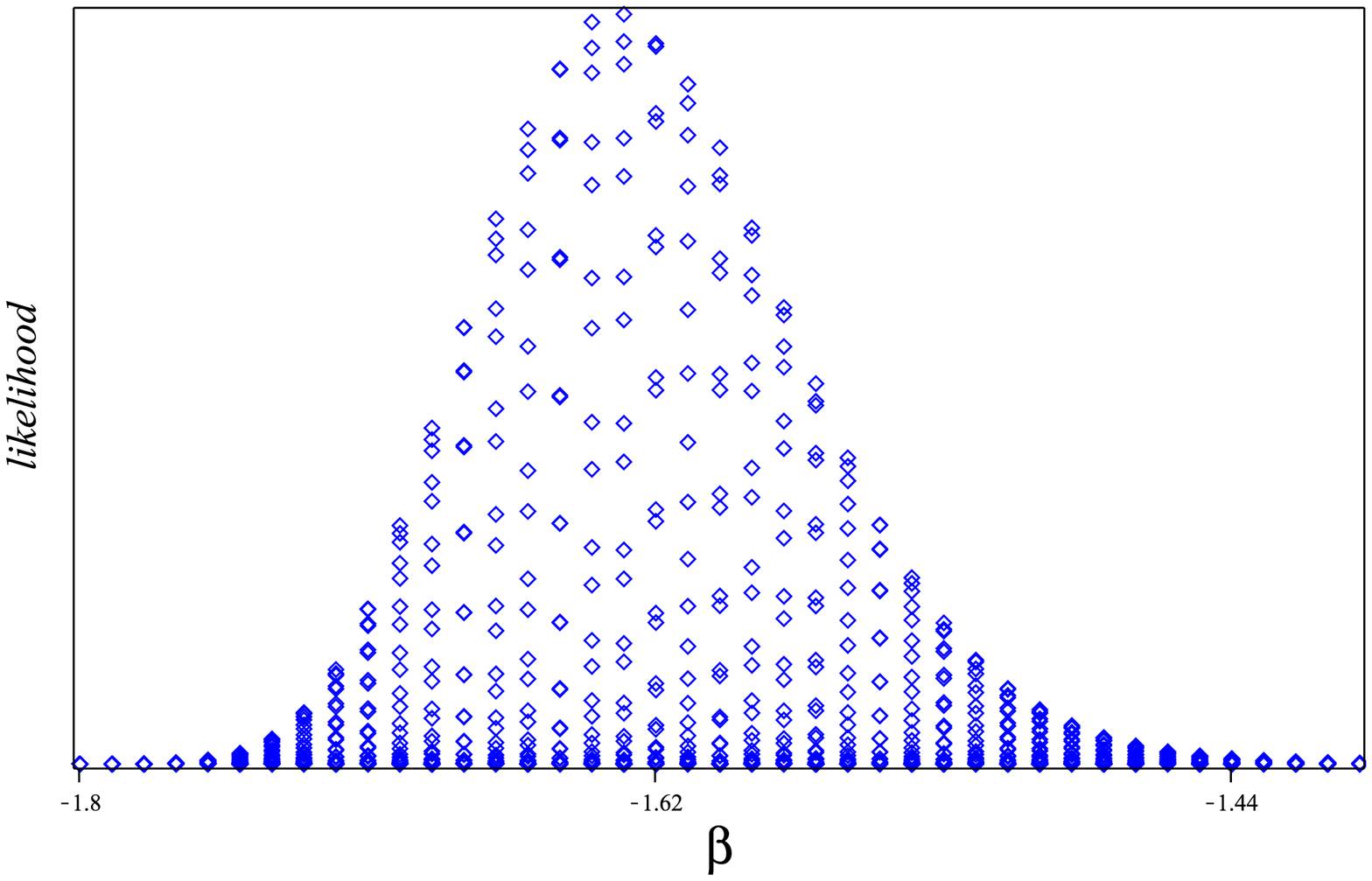}\hspace{0.1 cm}\includegraphics[scale=.28]{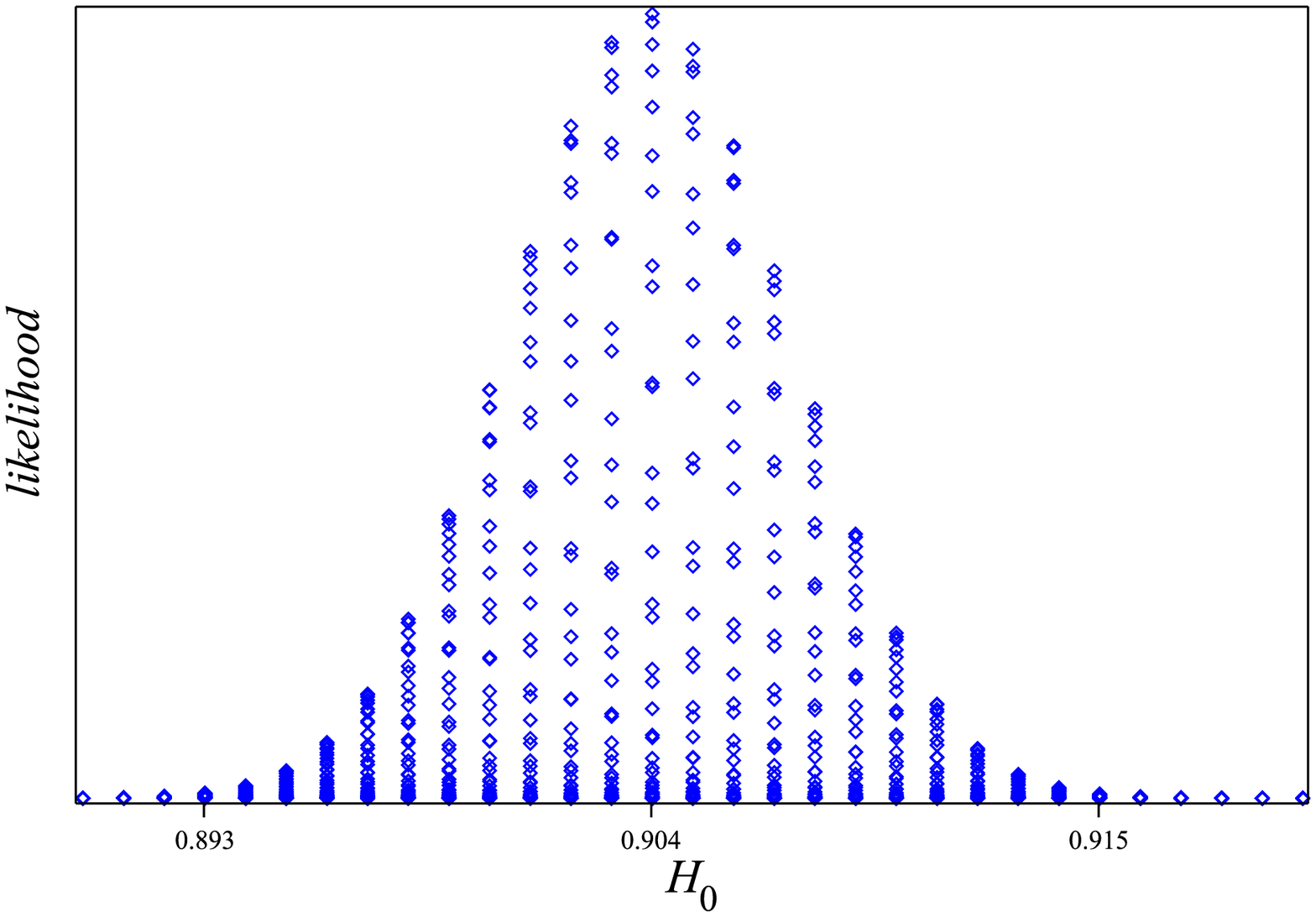}\hspace{0.1 cm}\\
Fig. 3:  The graph of 1-dim likelihood distribution for $\alpha$, $\beta$ and $H_{0}$ for $\gamma=0$
\end{tabular*}\\

\begin{tabular*}{2.5 cm}{cc}
\includegraphics[scale=.3]{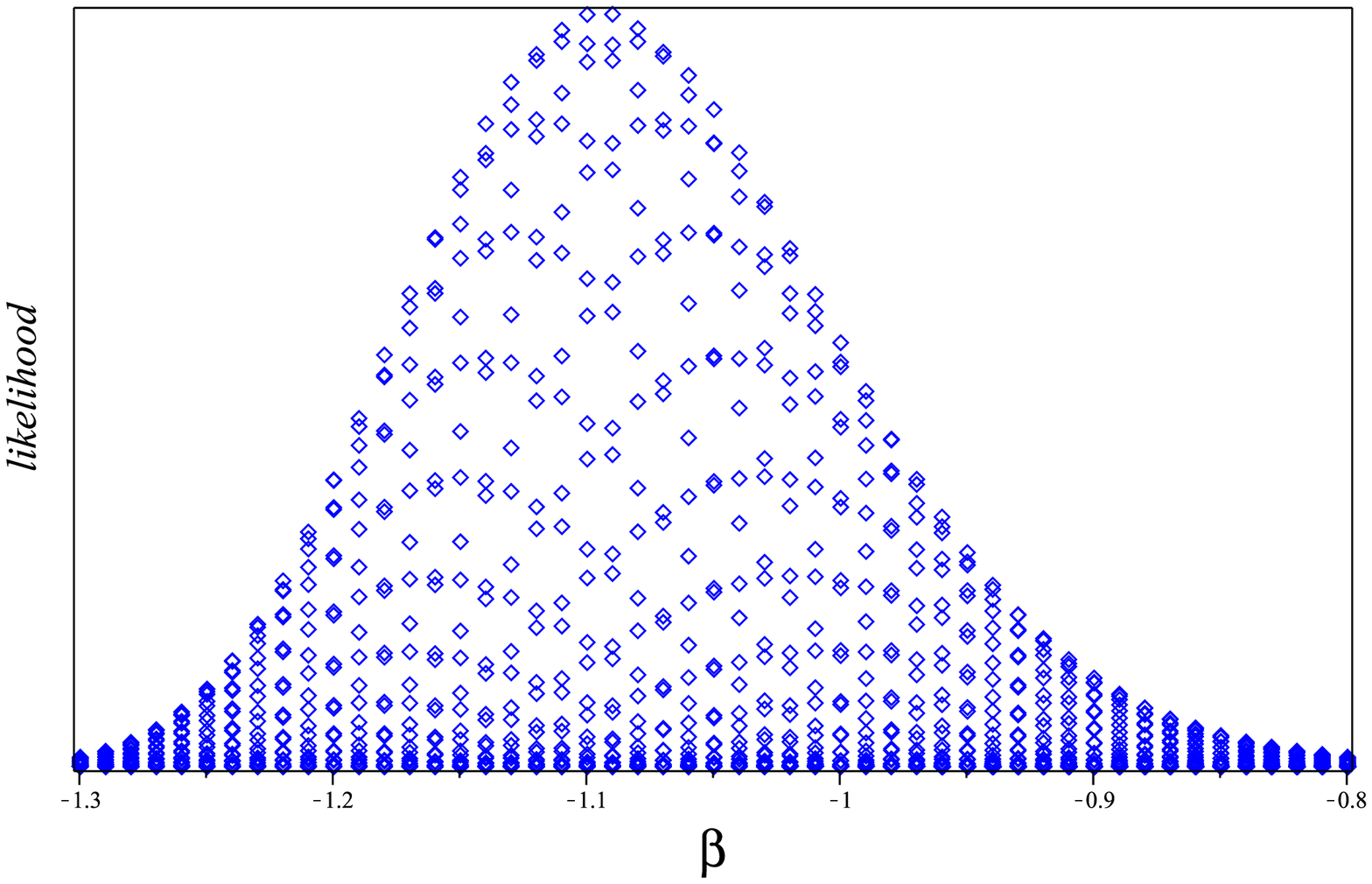}\hspace{0.1 cm}\includegraphics[scale=.26]{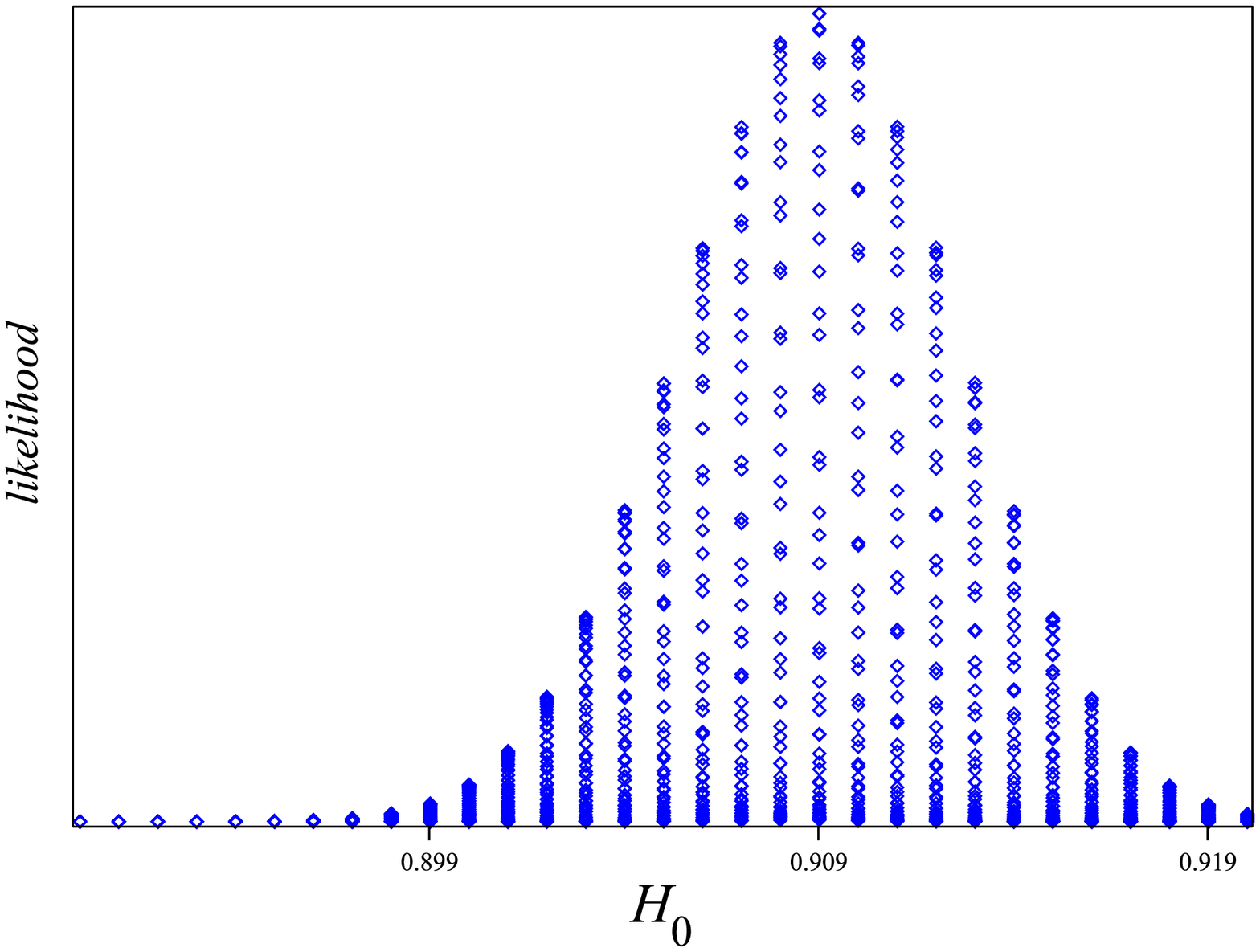}\hspace{0.1 cm}\\
Fig. 4:  The graph of 1-dim likelihood distribution for $\beta$ and $H_{0}$ for $\gamma=\frac{1}{3}$
\end{tabular*}\\

The distance modulus, $\mu(z)$, plotted in Fig. 5, in both cases $\gamma = 0, 1/3$ are best fitted with the SNe Ia observational data for the model parameters and initial conditions using $\chi^2$ method.

\begin{tabular*}{2.5 cm}{cc}
\includegraphics[scale=.45]{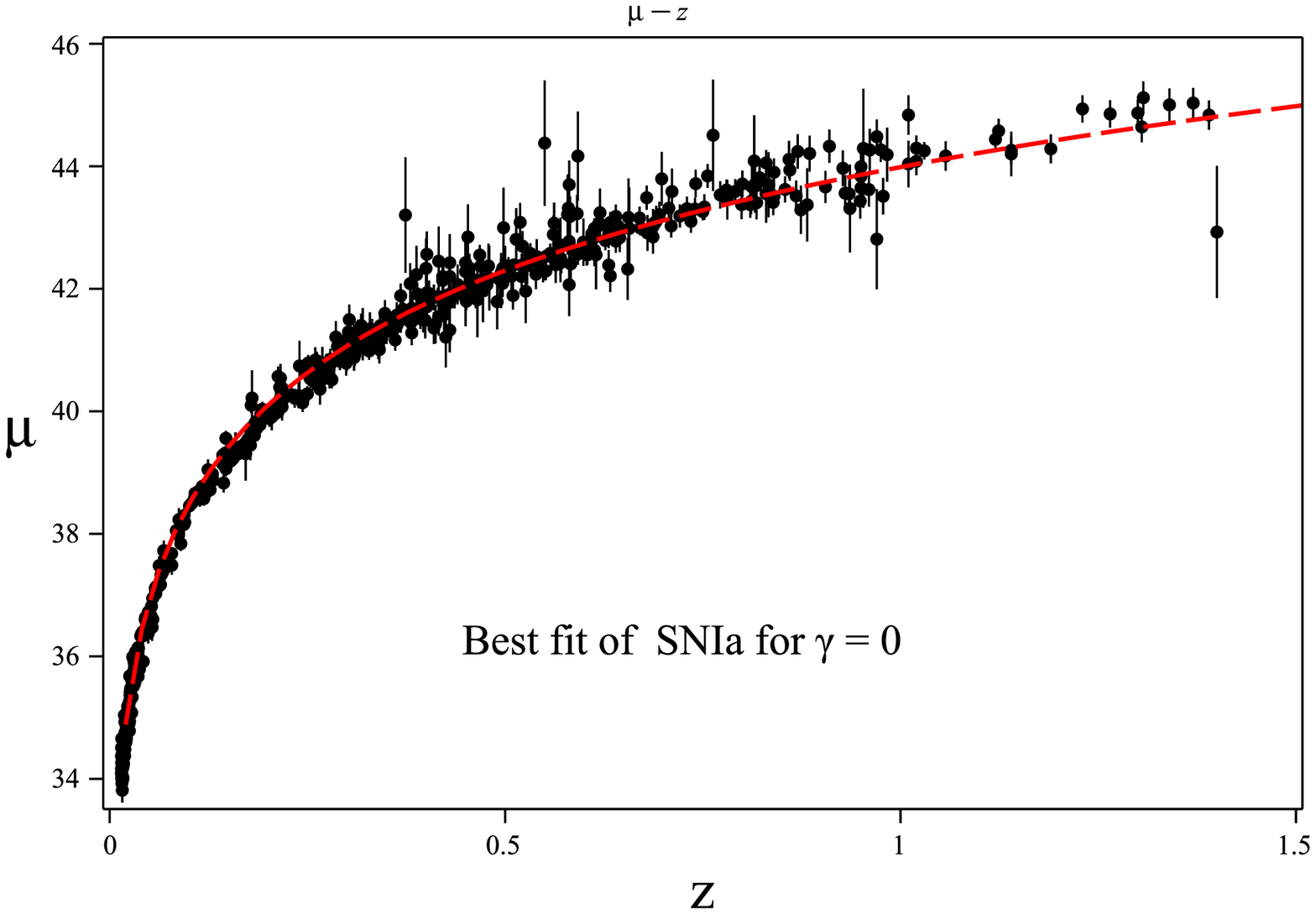}\hspace{0.1 cm}\includegraphics[scale=.45]{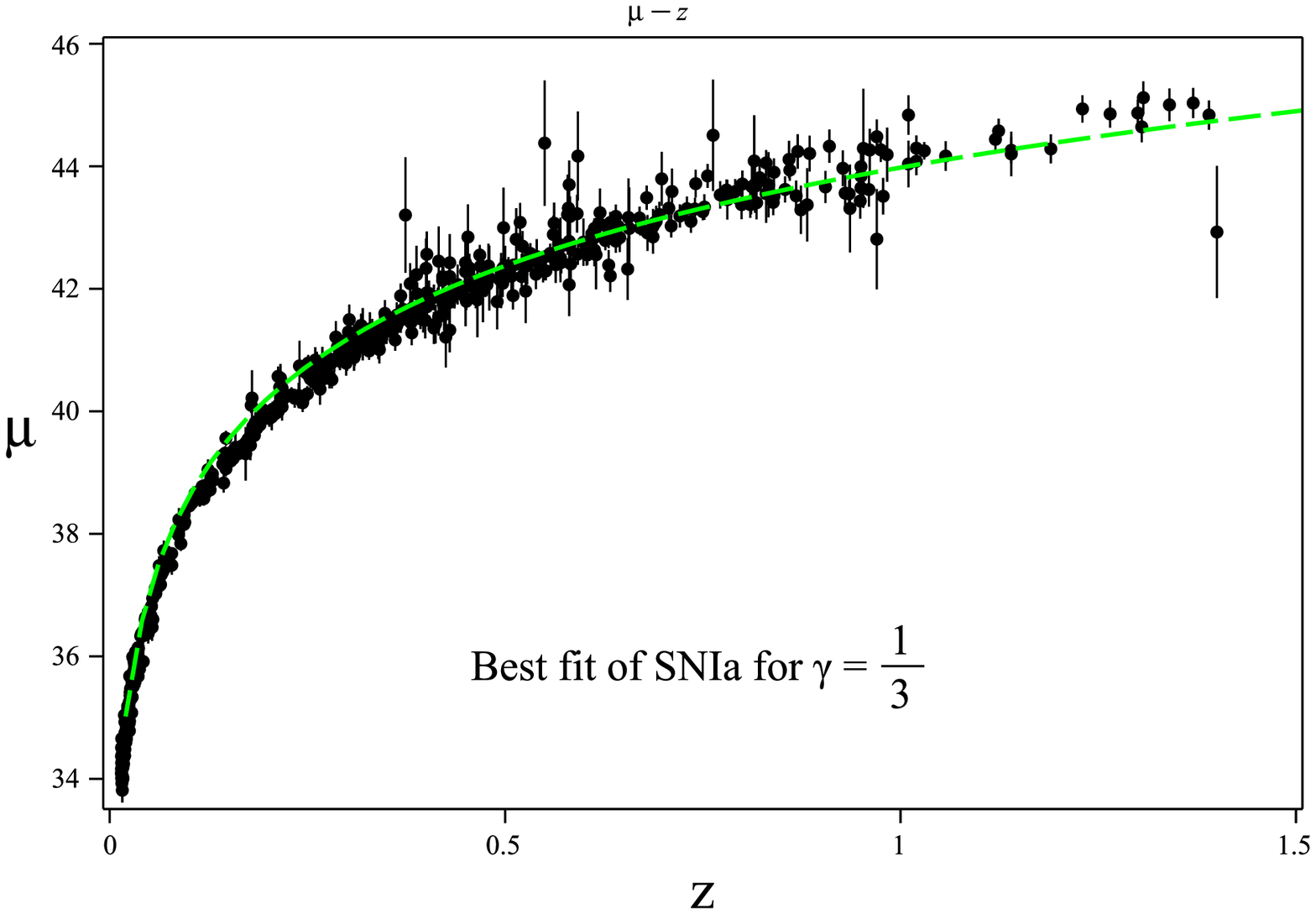}\hspace{0.1 cm}\\
Fig. 5: The best-fitted distance modulus $\mu(z)$ plotted as function of redshift for \\left) $\gamma=0$ right) $\gamma=\frac{1}{3}$
\end{tabular*}\\

In the following we investigate the stability of the model
with respect to the best fitted model parameters and initial conditions for the two specific choices of the EoS parameter for the matter in the universe, i.e. $\gamma=0$ and  $\gamma=1/3$.

\subsection{Stability of the best fitted critical points and phase space}

Solving the stability equations for the best fitted model parameters we find fixed points and the stability properties as illustrated in tables IV and V, for $\gamma=0$ and $\gamma=1/3$ respectively.\\

\begin{table}[ht]
\caption{Best-fitted fixed points for $\gamma=0$} 
\centering 
\begin{tabular}{c c c c c } 
\hline\hline 
points  & $(x, y)  $  & Stability \\
\hline 
 FP1&(0, 0) & unstable
 \\
 \hline
 FP2 & (0, 0.7) & stable
 \\
\hline 
 FP3&$(0, -0.7) $ & stable
 \\
\end{tabular}
\label{table:1} 
\end{table}\

\begin{table}[ht]
\caption{Best-fitted fixed points for $\gamma=\frac{1}{3}$} 
\centering 
\begin{tabular}{c c c c c } 
\hline\hline 
points  & $(x, y)  $  & Stability \\
\hline 
FP1&(0, 0) & unstable
 \\
 \hline
 FP2 & (0, 0.8) & stable
 \\
\hline 
 FP3&$(0, -0.8) $ & stable
 \\
\end{tabular}
\label{table:1} 
\end{table}\

From the above tables we see that, by best fitting the stability parameters, the critical points FP2 and FP3 are stable whereas FP1 is unstable in both $\gamma=1/3,0$. Thus, the number of critical points reduced from three to two. In addition, when best-fitting both the stability parameters and initial conditions, the best fitted trajectory for the system can be obtained which starts from an unstable critical point or infinity and approaches the stable critical point. In both graphs in Fig. 6), for $\gamma=0, 1/3$ all the best fitted trajectories for the stability parameters with different initial conditions leaving the unstable critical point P1 in the past in the phase plane is shown moving towards the stable critical point P2 and P3 in the future. Moreover, the best fitted trajectories ( in red color) for both stability parameters and initial conditions are shown approaching the critical point P3.

\begin{tabular*}{2.5 cm}{cc}
\includegraphics[scale=.45]{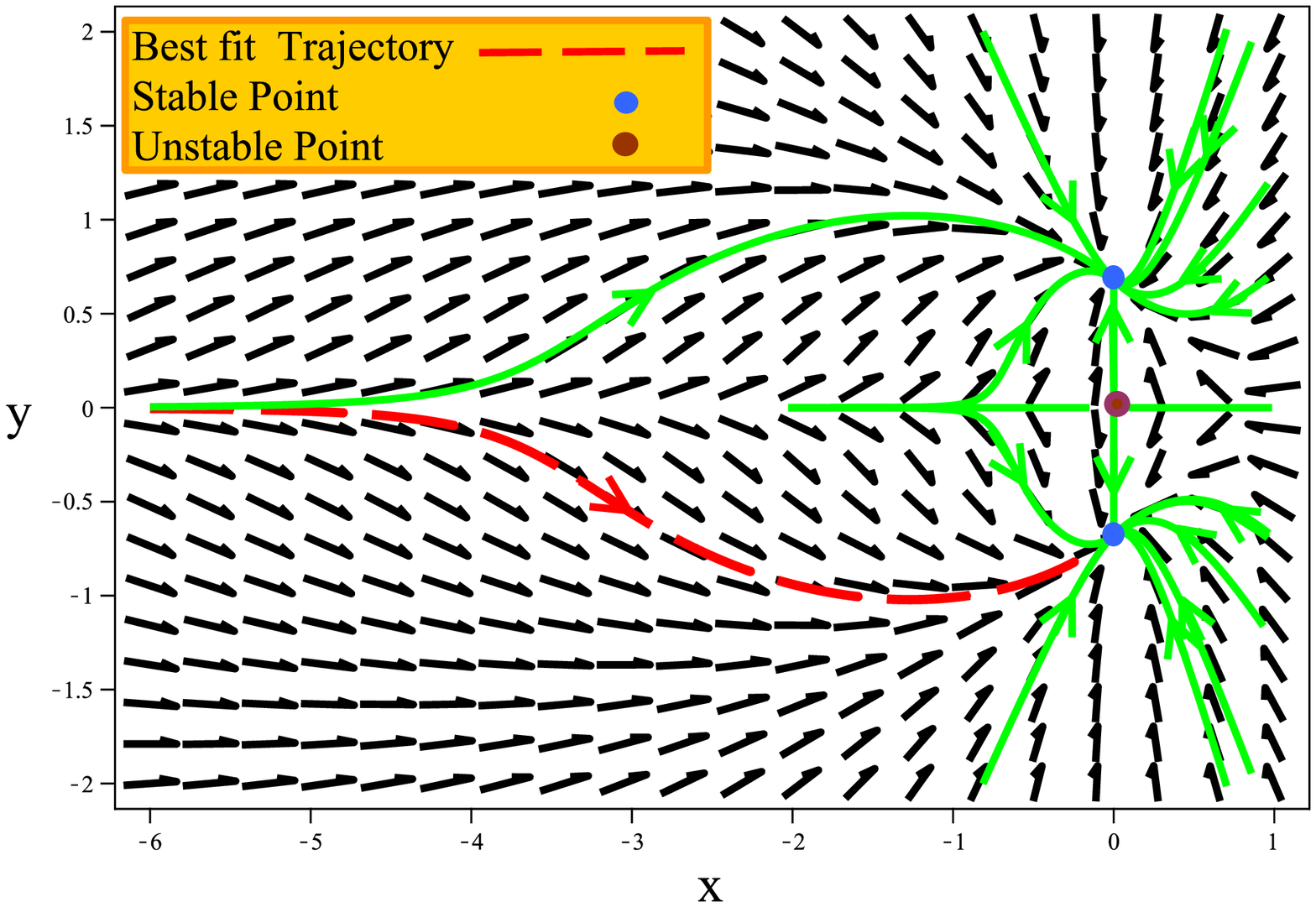}\hspace{0.1 cm}\includegraphics[scale=.45]{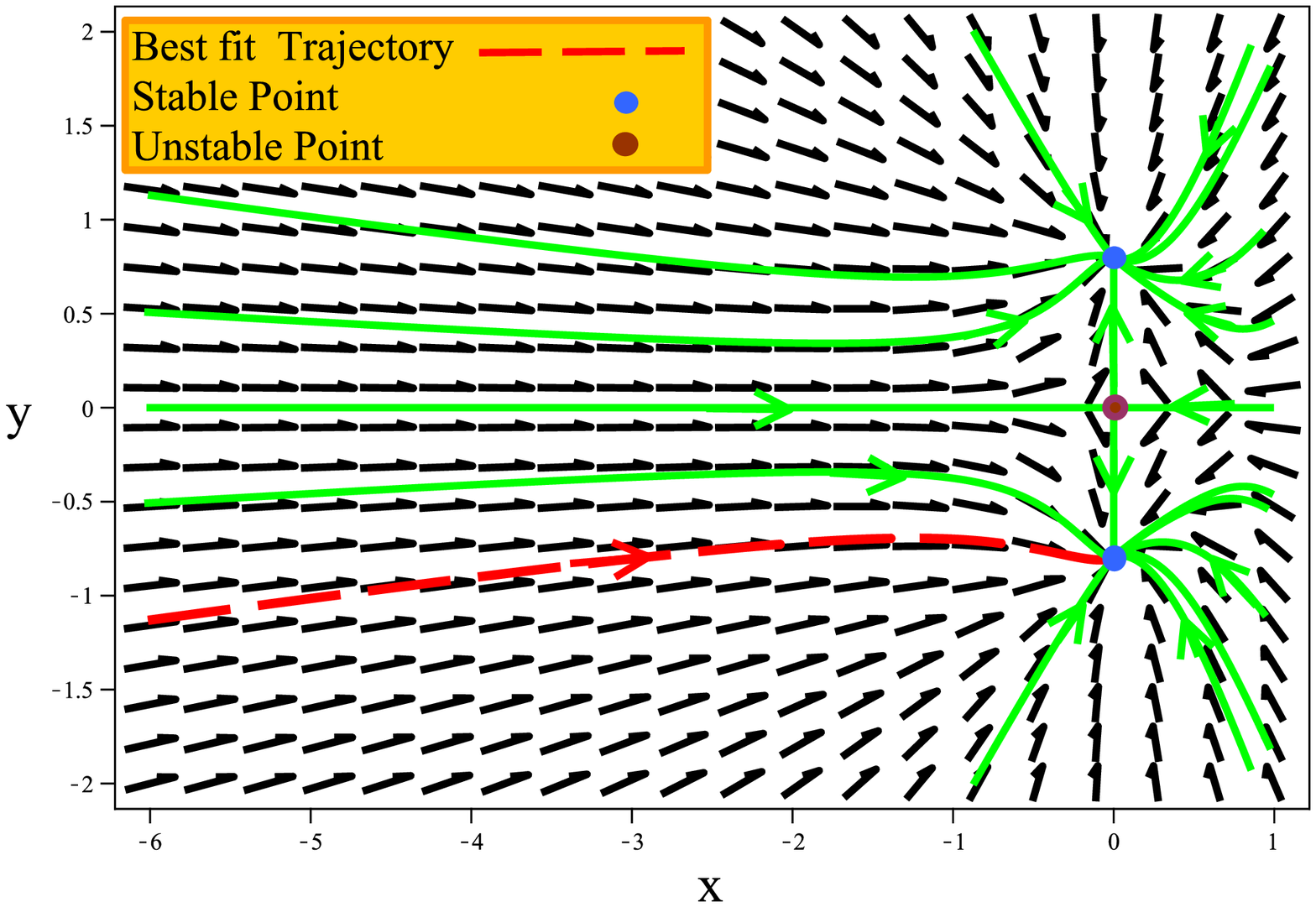}\hspace{0.1 cm}\\
Fig. 6: The phase plane for left) $\gamma=0$, right) $\gamma=\frac{1}{3}$ for best-fitted model parameter $ \alpha$ and $\beta$\\
\end{tabular*}\\

\section{Cosmological test}
To verify our model with the observation, in the following we perform two different kind of cosmological tests.

\subsection{Cosmological parameters}
In order to understand the behavior of the universe and its dynamics we need to study the cosmological parameters. We have best fitted our model with the current observational data by the distance modulus test. The cosmological parameters analytically and/or numerically have been investigated by many authors for variety of cosmological models. Applying stability analysis and simultaneously best fitting the model with the observational data gives us a better understanding of the critical points and the dynamics of these parameters. The effective EoS parameter, deceleration parameter and statefinders are given in terms of new  dynamical variables in equations (\ref{eff}-\ref{s}). In tables VI and VII, the properties of the best-fitted critical points for $\gamma=0, 1/3$ are shown.\\

\begin{table}[hb]
\caption{Properties of the best-fitted fixed points for $\gamma=0$ } 
\centering 
\begin{tabular}{c c c c c c  } 
\hline\hline 
points   & q & $\omega_{eff}$ & r & s & acceleration \\ [3ex] 
\hline 
FP1  & 1/2 & 0 & $1$&$1$& No\\ 
FP2  & $-0.2$ & $-0.47$ & $-0.11$&$0.54$& yes \\
FP3 & $-0.2$ & $-0.47$ & $-0.11$&$0.54$& yes \\
 [1ex] 
\hline 
\end{tabular}
\label{table:2} 
\end{table}

\begin{table}[hb]
\caption{Properties of the best-fitted fixed points for $\gamma=\frac{1}{3}$ } 
\centering 
\begin{tabular}{c c c c c c  } 
\hline\hline 
points   & q & $\omega_{eff}$ & r & s & acceleration \\ [3ex] 
\hline 
FP1  & 1/2 & 0 & $1$&$1$& No\\ 
FP2  & $-0.45$ & $-0.63$ & $-0.02$&$0.37$& yes \\
FP3  & $-0.45$ & $-0.63$ & $-0.02$&$0.37$& yes \\
 [1ex] 
\hline 
\end{tabular}
\label{table:2} 
\end{table}\

In Fig. 7 and 8, the effective EoS parameters and deceleration parameter are shown in both cases, $\gamma=0, 1/3$.  All the trajectories shown in these graphs are best fitted for the stability parameters $\alpha$ and $\beta$. In addition, the trajectory with red color are best fitted for both stability parameters and also initial conditions.  From Fig. 7 one observes that for $\gamma=0, 1/3$ the phantom crossing do not occur in the past and future. In Fig. 7)left and table VI, for $\gamma=0$ it shows that if both stability parameters and initial conditions are best-fitted, the universe starts from the unstable state at in the past with $\omega_{eff}=0$ and approaches the stable state in future with $\omega_{eff}=-0.47$ (FP3). The best-fitted current value for the effective EoS parameter is about $-5.5$.

 From Fig. 7)right and table VII, for $\gamma=1/3$ we see that the universe start from unstable state in the past with $\omega_{eff}\rightarrow -\infty$ and approaches the stable state in future with $\omega_{eff}=-0.43$ (FP3). The best-fitted current value for the effective EoS parameter is about $-6.2$.

 From Fig.8 for $\gamma=0$, we see that, the universe start from unstable state with positive deceleration parameter and approaches the accelerating stable state in future. It is interesting to see from the graph that for the best fitted stability parameters the universe is currently decelerating whereas for the best fitted stability parameters and initial conditions the universe is currently in accelerating state which is in compatible with the observational data.
 One see that from Fig.8 for $\gamma=1/3$, the universe starts from unstable state in the past with $q\rightarrow -\infty$ which is not compatible with observation.\\

\begin{tabular*}{2.5 cm}{cc}
\includegraphics[scale=.4]{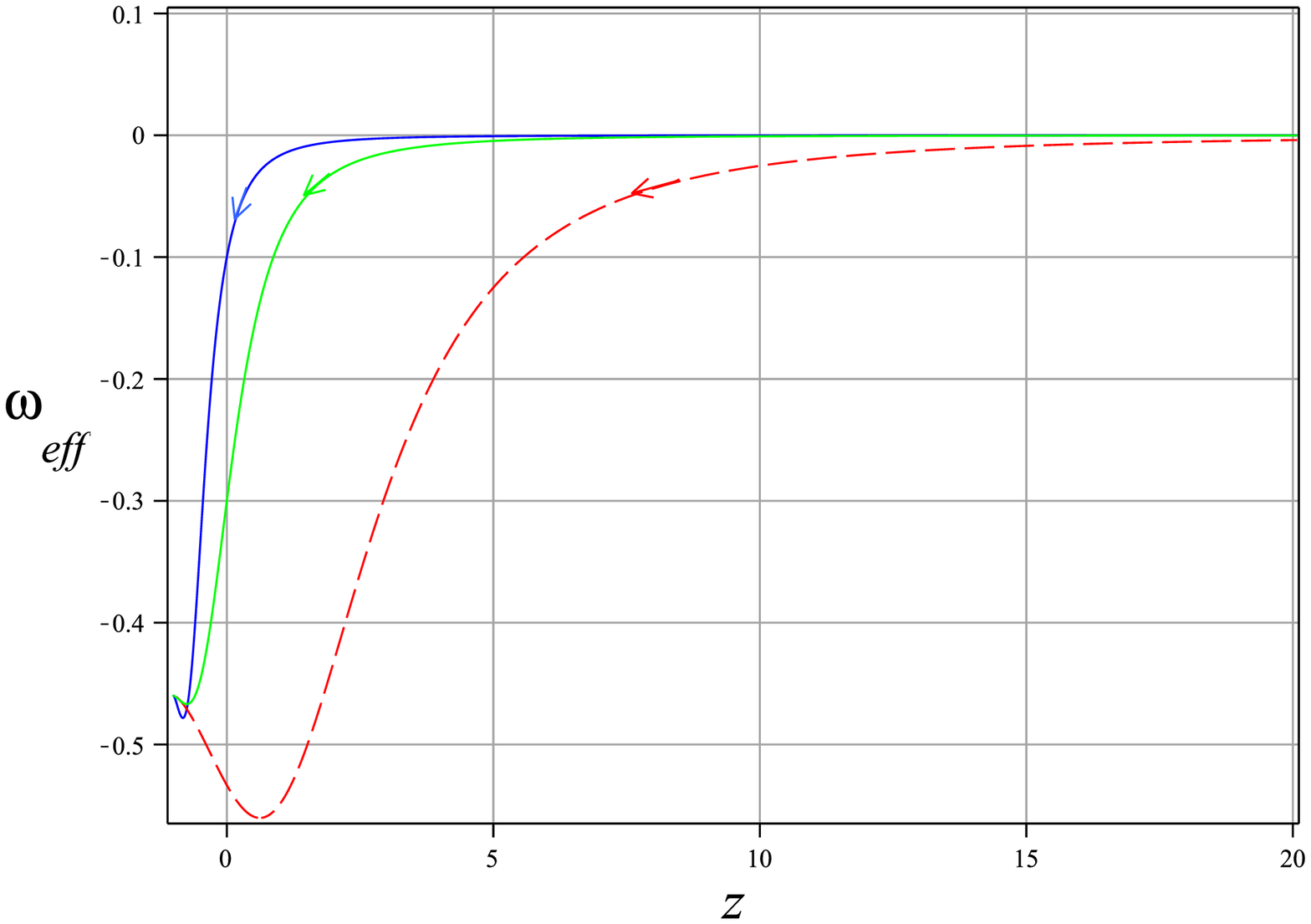}\hspace{0.1 cm}\includegraphics[scale=.4]{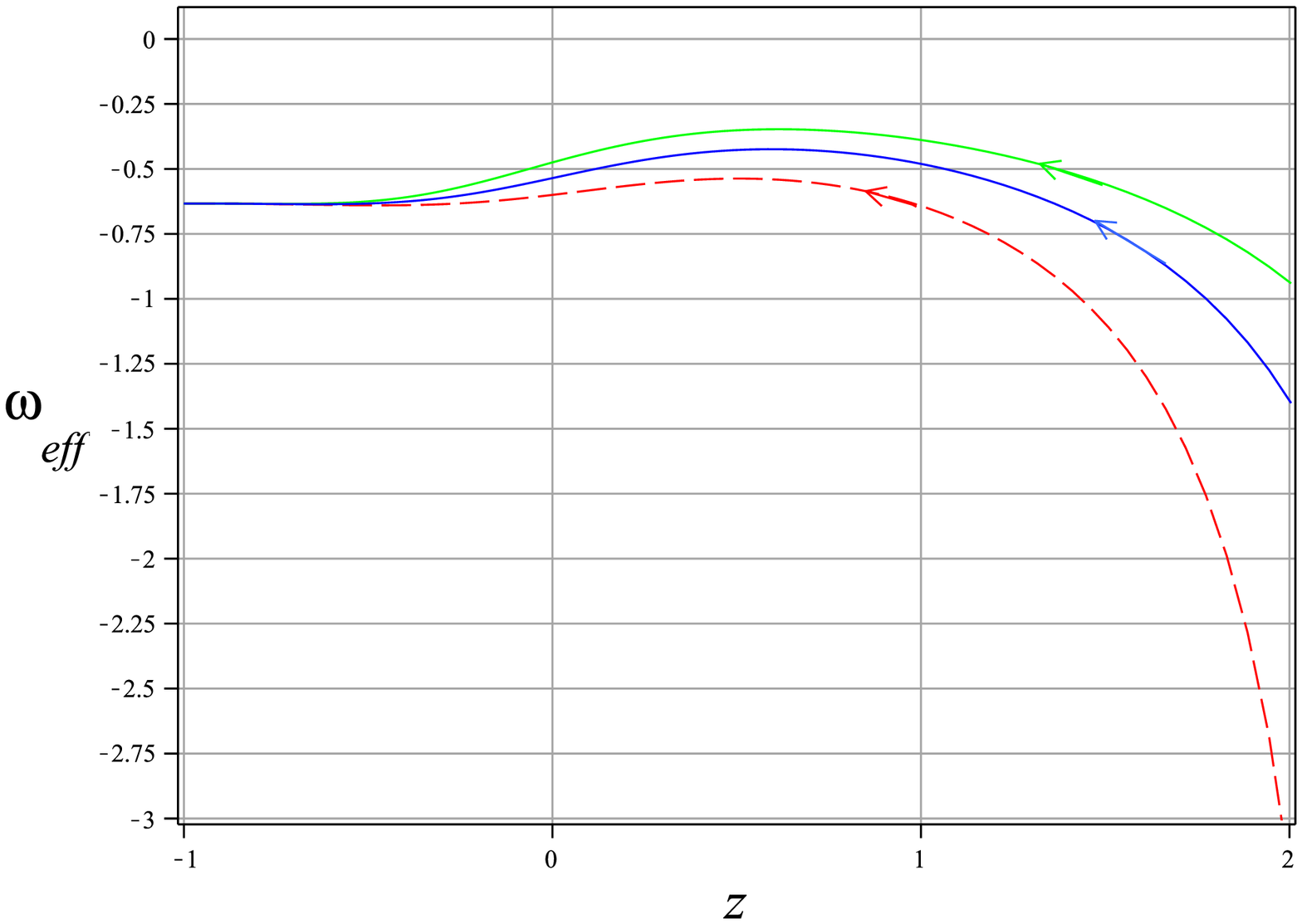}\hspace{0.1 cm}\\
Fig.7: The graph of best-fitted effective EoS parameter, $\omega_{eff}$ for left) $\gamma=0$, right) $\gamma=\frac{1}{3}$.\\ The red trajectory is for the both best-fitted model parameters and I.C.s
\end{tabular*}\\
\begin{tabular*}{2.5 cm}{cc}
\includegraphics[scale=.4]{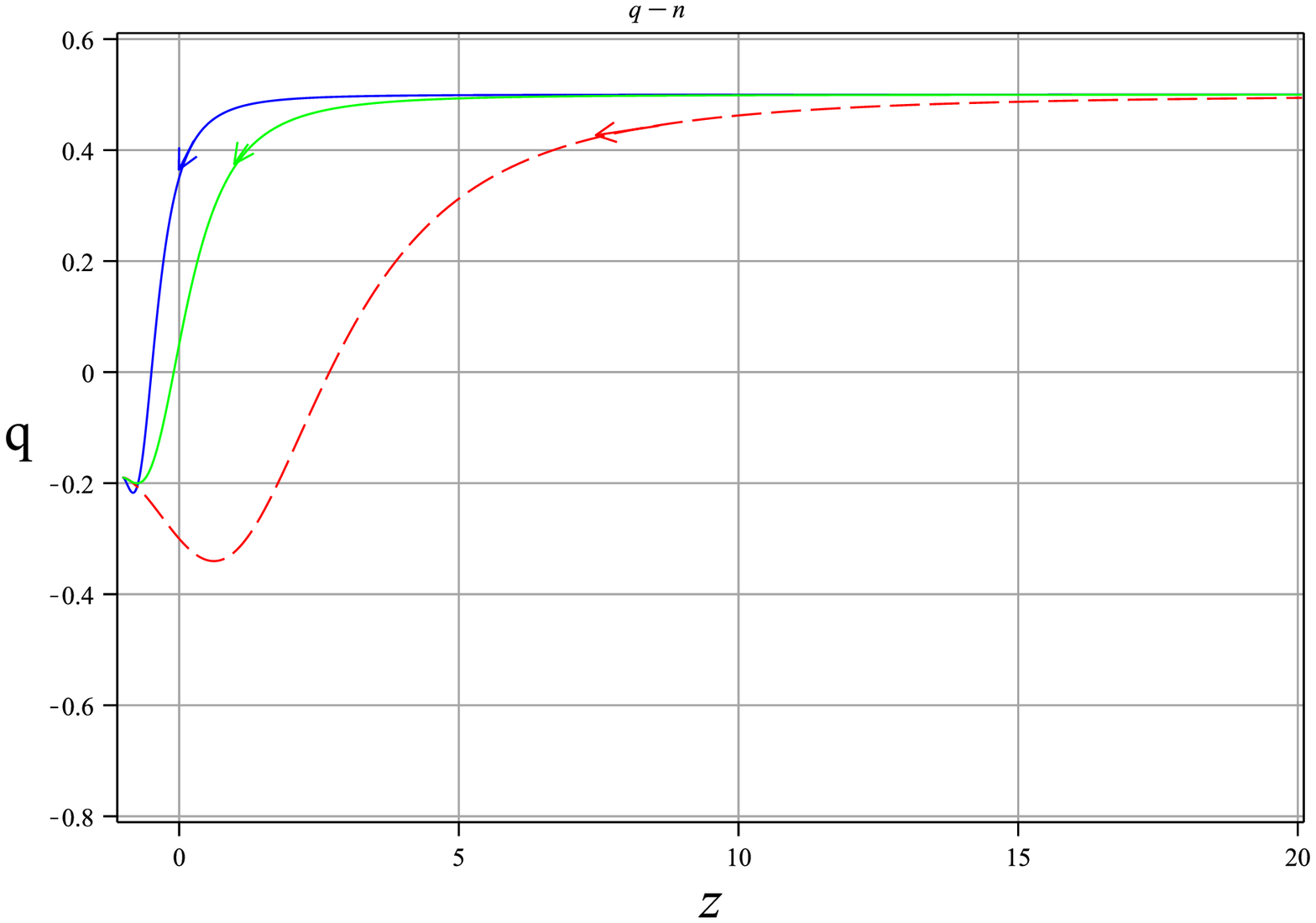}\hspace{0.1 cm}\includegraphics[scale=.4]{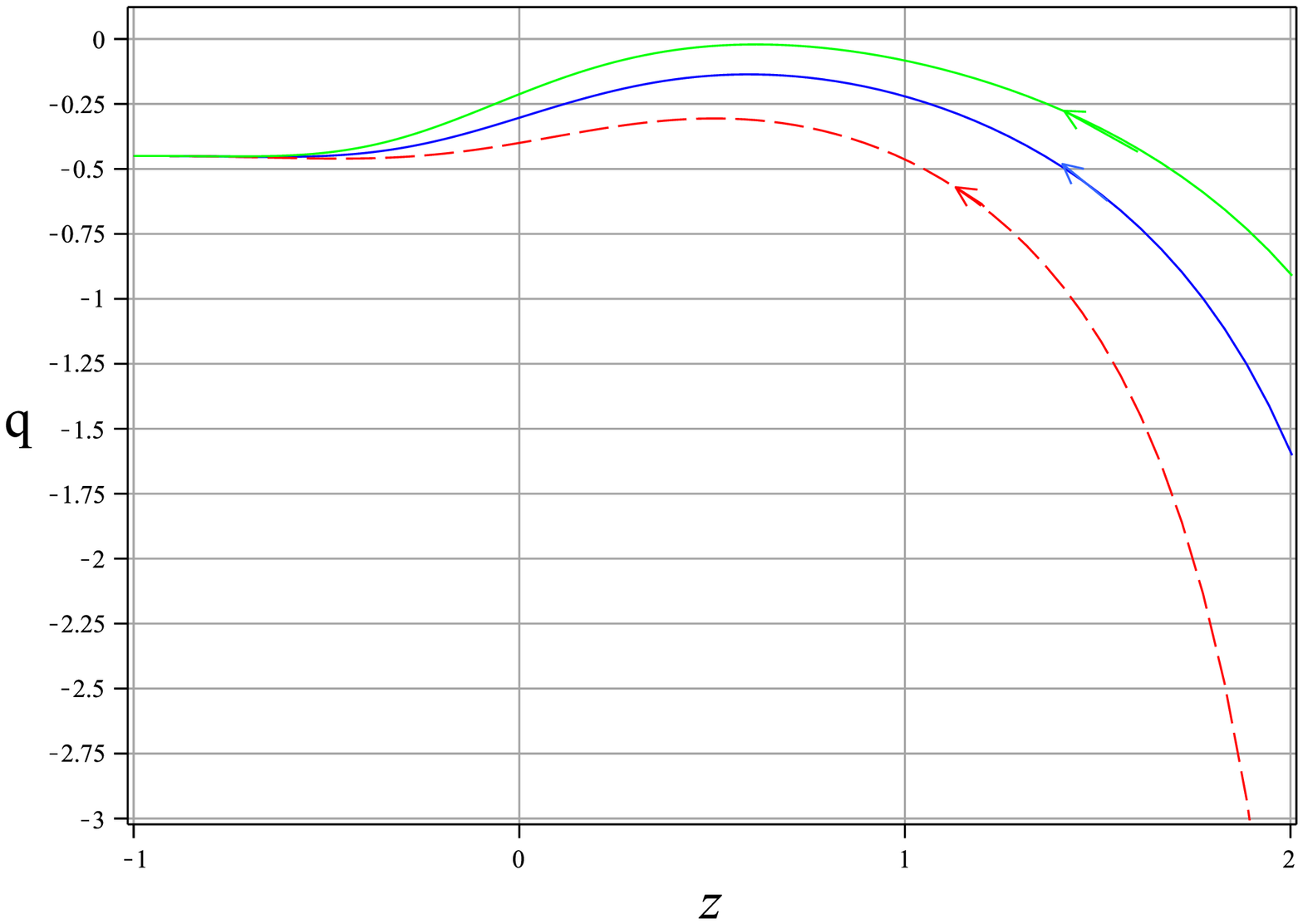}\hspace{0.1 cm}\\
Fig.8: The graph of best-fitted deceleration parameter, $q$ for left) $\gamma=0$, right) $\gamma=\frac{1}{3}$.\\ The red trajectory is for the both best-fitted model parameters and best I.C.s
\end{tabular*}\\

Figs.9-13 show the statefinder diagrams $\{s,q\}$ and $\{r,q\}$ and their evolutionary
trajectories. As above, in all these graphs, the statefinder trajectories are best-fitted for the stability parameters, whereas the trajectories with red color are also best fitted for initial conditions.

From the graph of the statefinder $\{r,q\}$, $\{s,q\}$ and $\{r,s\}$ for $\gamma=0$ we see that all the best-fitted trajectories start from the standard cold dark energy (SCDM) in the past which is an
unstable critical point and tend their evolution to the stable state in the
future. The current values of the trajectories are also shown and can be compared with the
position of SCDM. However the result is not promising for the case $\gamma=1/3$ as the trajectories start from not a physically known state in the past.\\

\begin{tabular*}{2.5 cm}{cc}
\includegraphics[scale=.4]{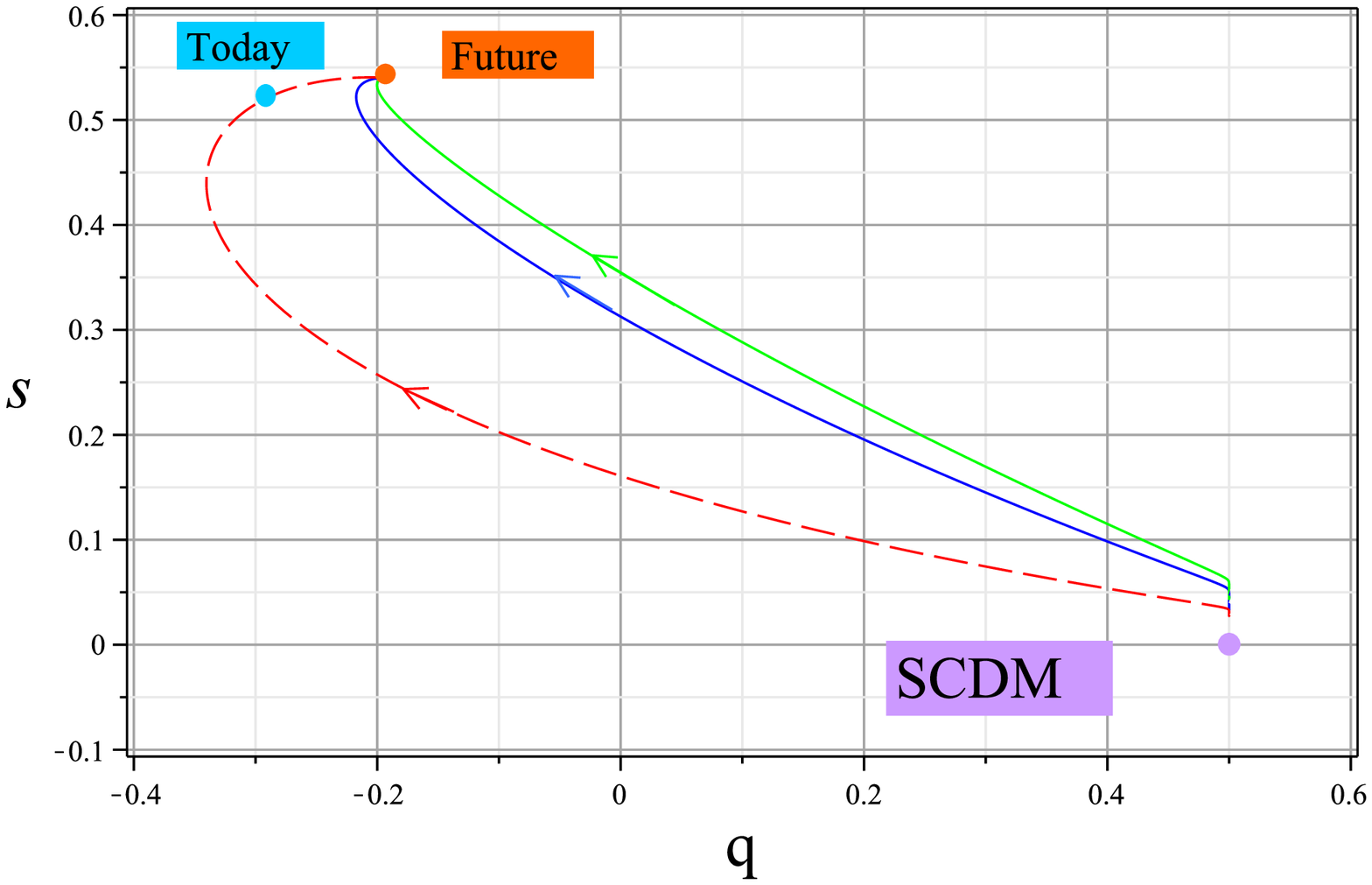}\hspace{0.1 cm}\includegraphics[scale=.4]{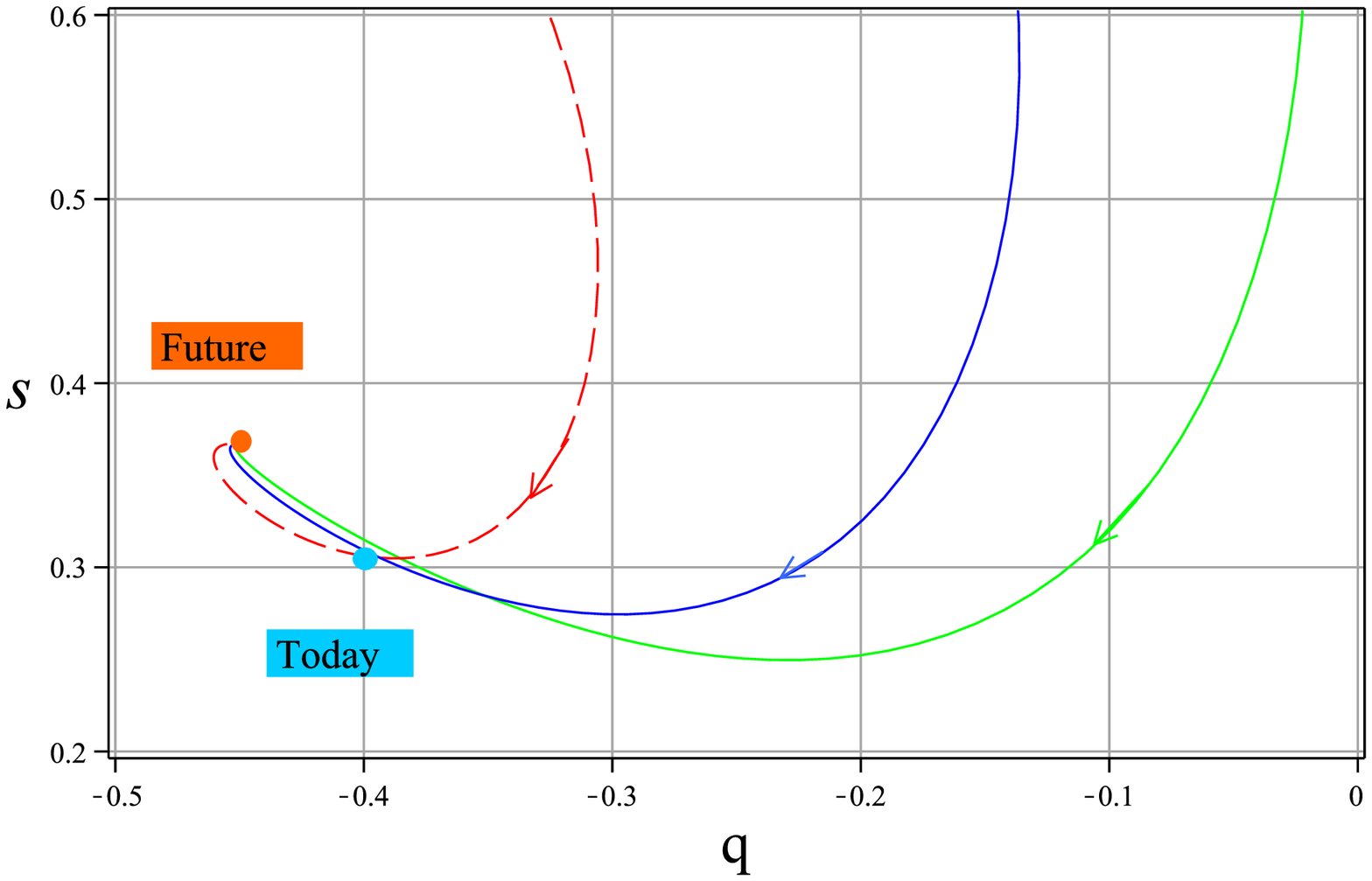}\hspace{0.1 cm}\\
Fig.9:  The graph of the best-fitted statefinder  $\{s,q\}$ for left) $\gamma=0$, right) $\gamma=\frac{1}{3}$.\\ The red trajectory is for the both best-fitted model parameters and best I.C.s
\end{tabular*}\\
\begin{tabular*}{2.5 cm}{cc}
\includegraphics[scale=.4]{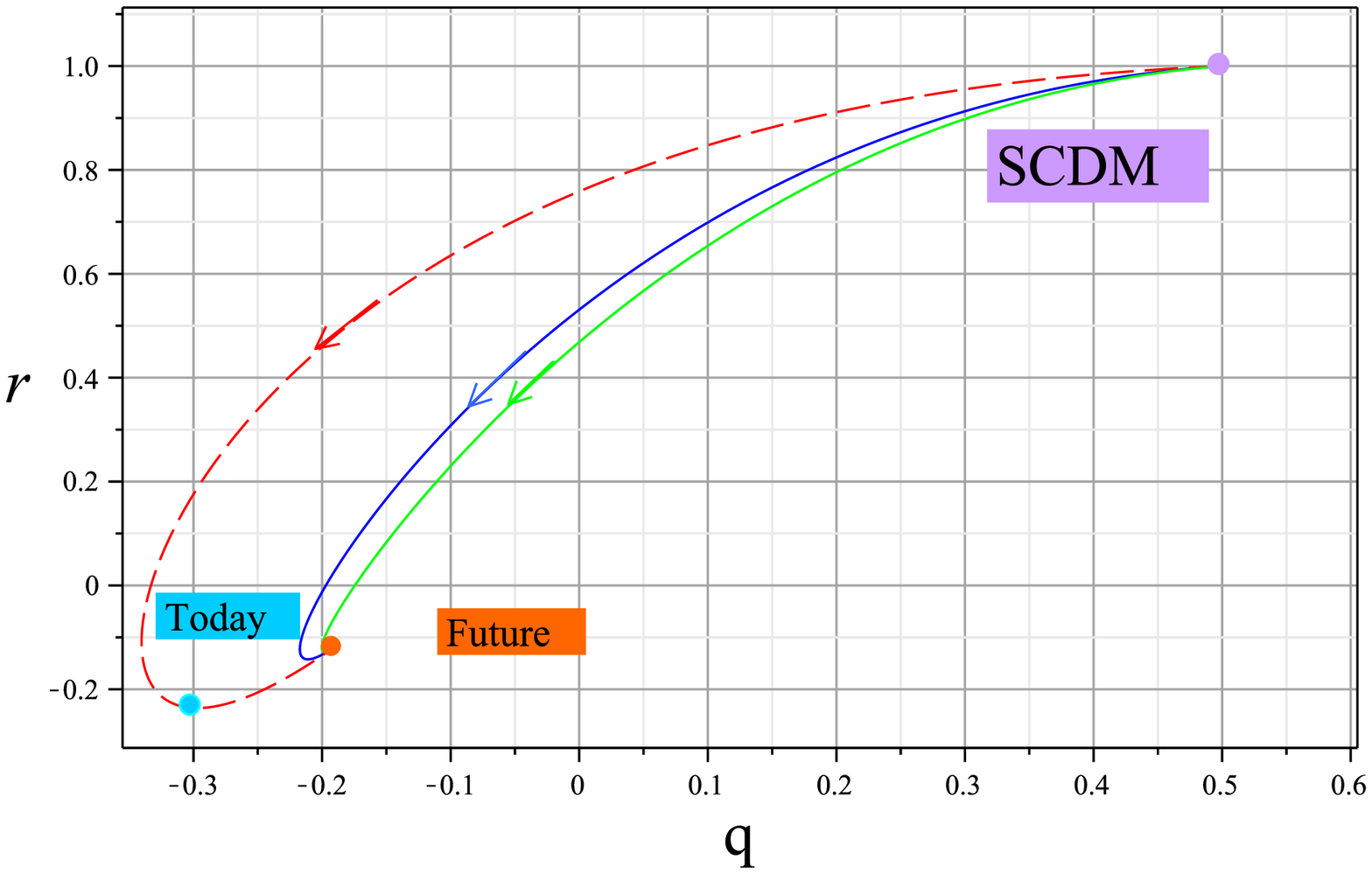}\hspace{0.1 cm}\includegraphics[scale=.4]{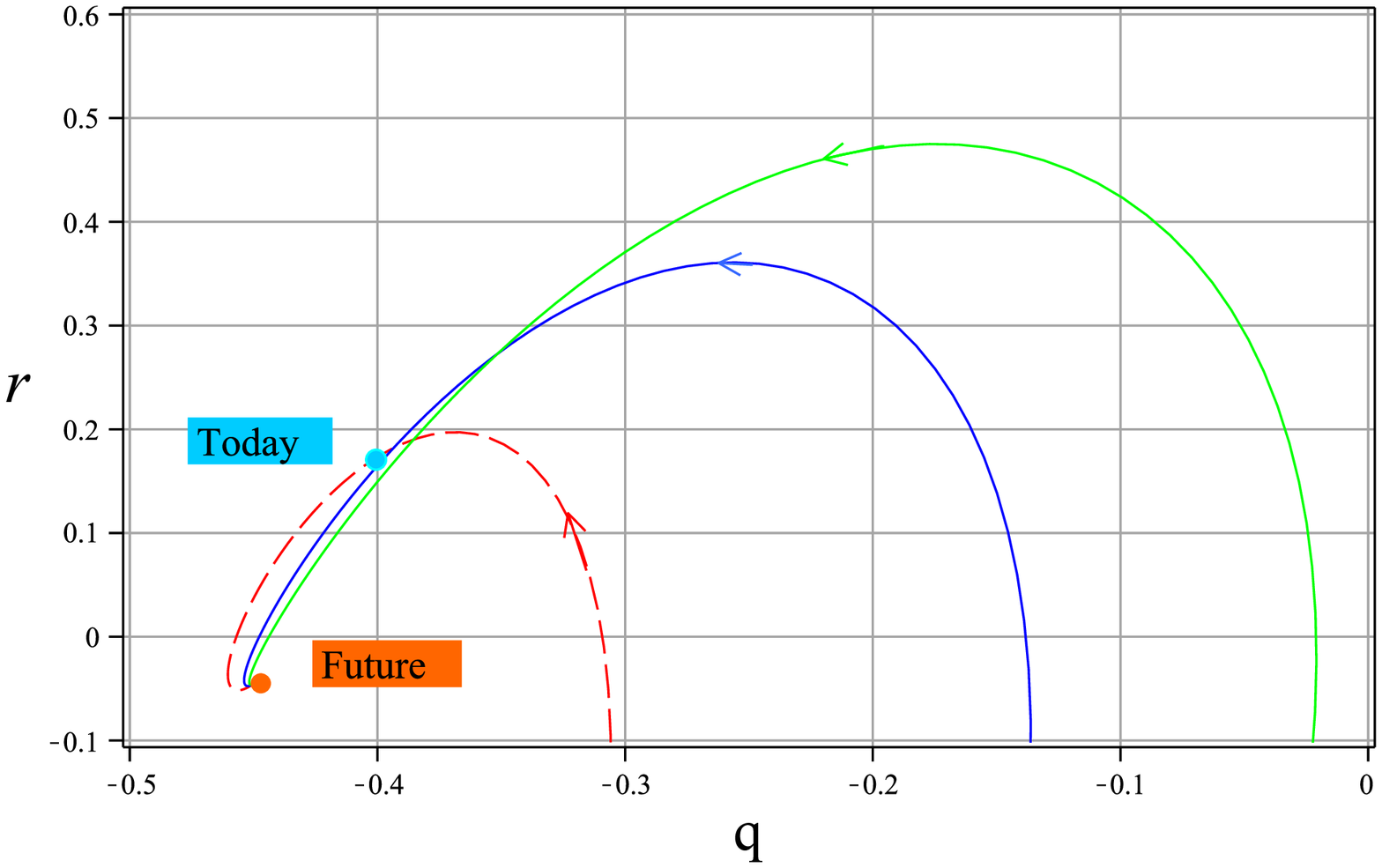}\hspace{0.1 cm}\\
Fig.10:  The graph of the best-fitted statefinder  $\{r,q\}$ for left) $\gamma=0$, right) $\gamma=\frac{1}{3}$.\\ The red trajectory is for the both best-fitted model parameters and best I.C.s
\end{tabular*}\\
\begin{tabular*}{2.5 cm}{cc}
\includegraphics[scale=.4]{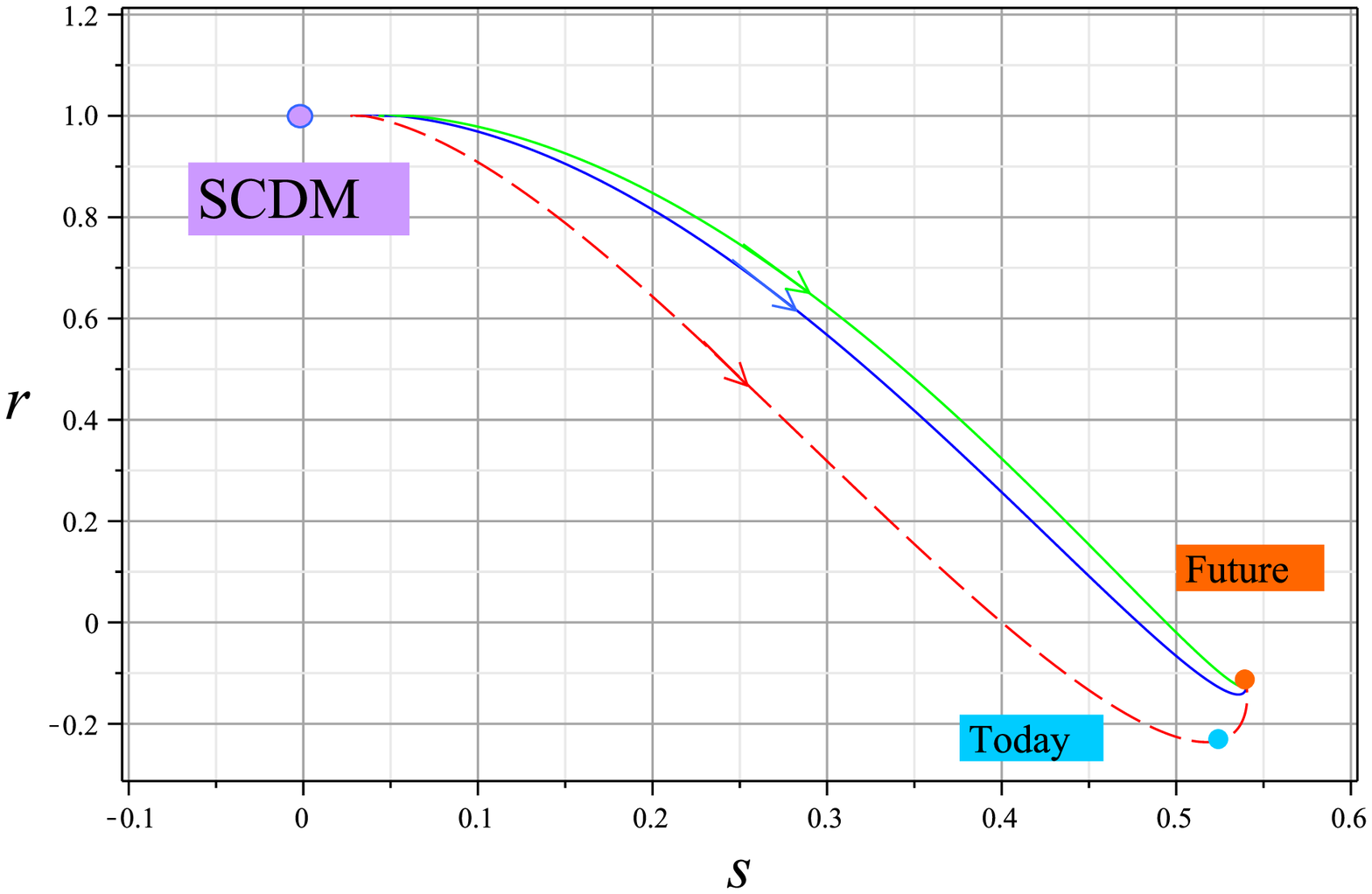}\hspace{0.1 cm}\includegraphics[scale=.4]{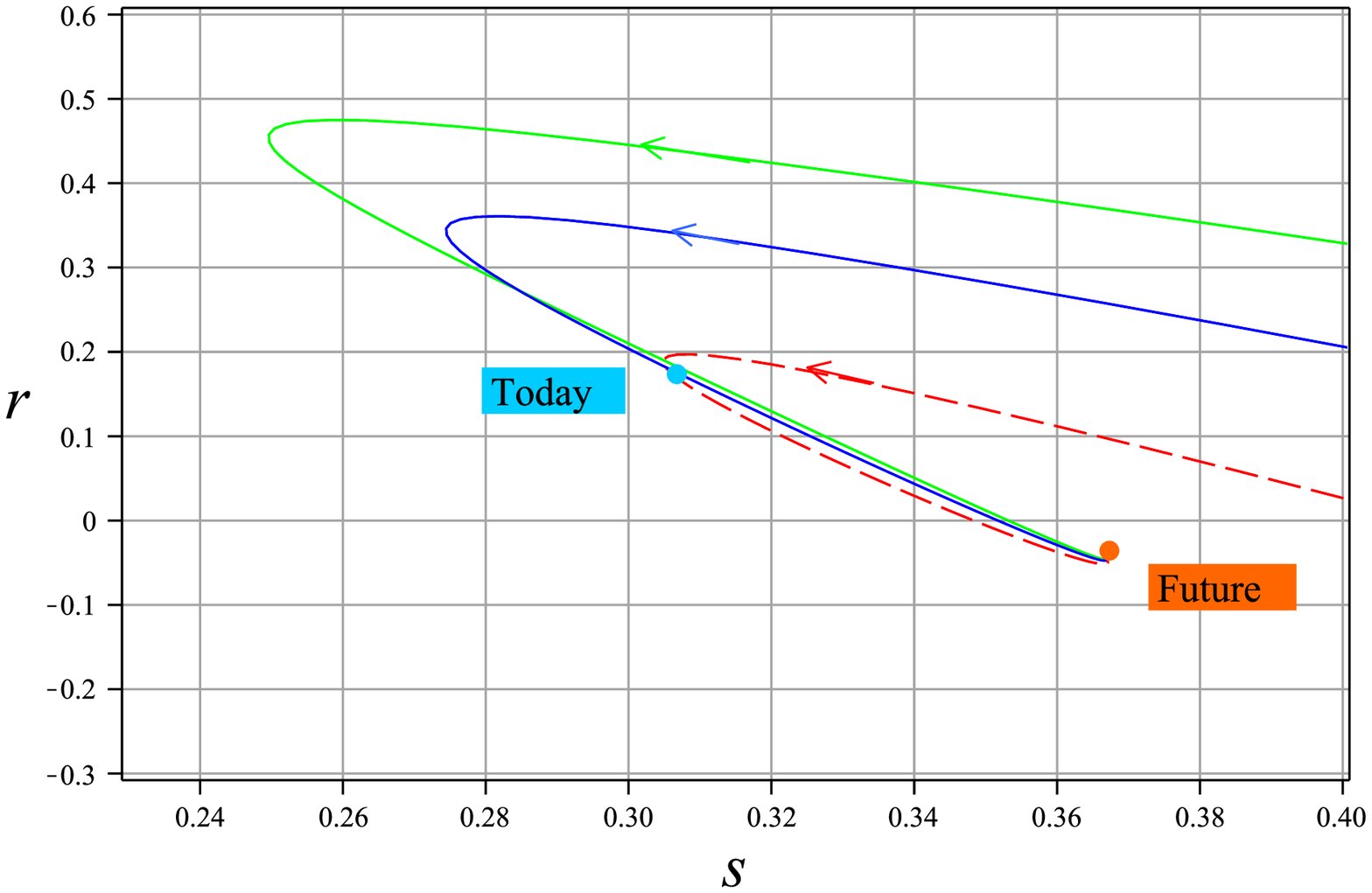}\hspace{0.1 cm}\\
Fig.11:  The graph of the best-fitted statefinder  $\{r,s\}$ for left) $\gamma=0$, right) $\gamma=\frac{1}{3}$.\\ The red trajectory is for the both best-fitted model parameters and best I.C.s
\end{tabular*}\\
\begin{tabular*}{2.5 cm}{cc}
\includegraphics[scale=.4]{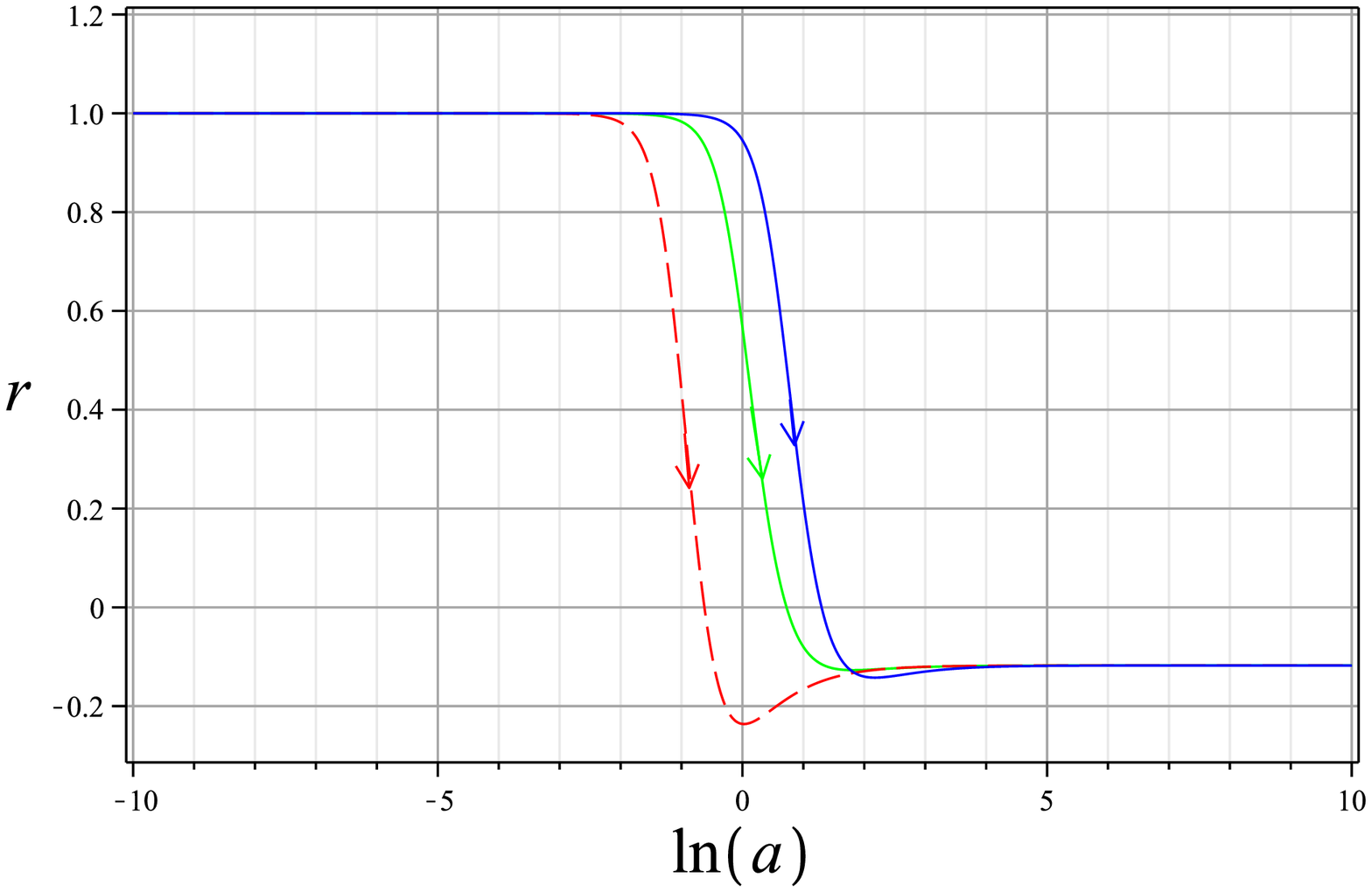}\hspace{0.1 cm}\includegraphics[scale=.4]{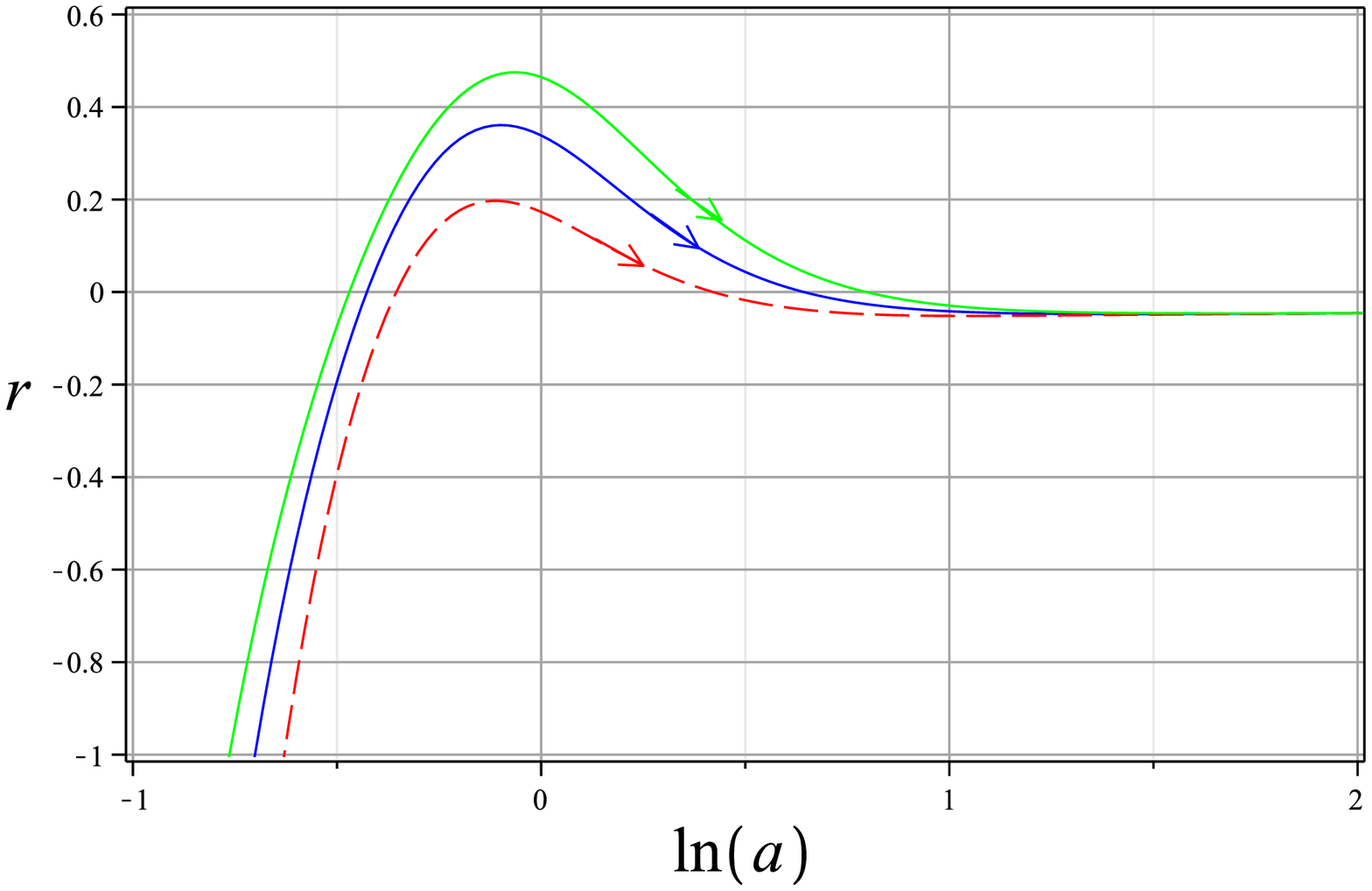}\hspace{0.1 cm}\\
Fig.12:  The graph of the best-fitted statefinder  $ ({r,\ln (a))}$ for left) $\gamma=0$, right) $\gamma=\frac{1}{3}$.\\ The red trajectory is for the both best-fitted model parameters and best I.C.s
\end{tabular*}\\
\begin{tabular*}{2.5 cm}{cc}
\includegraphics[scale=.4]{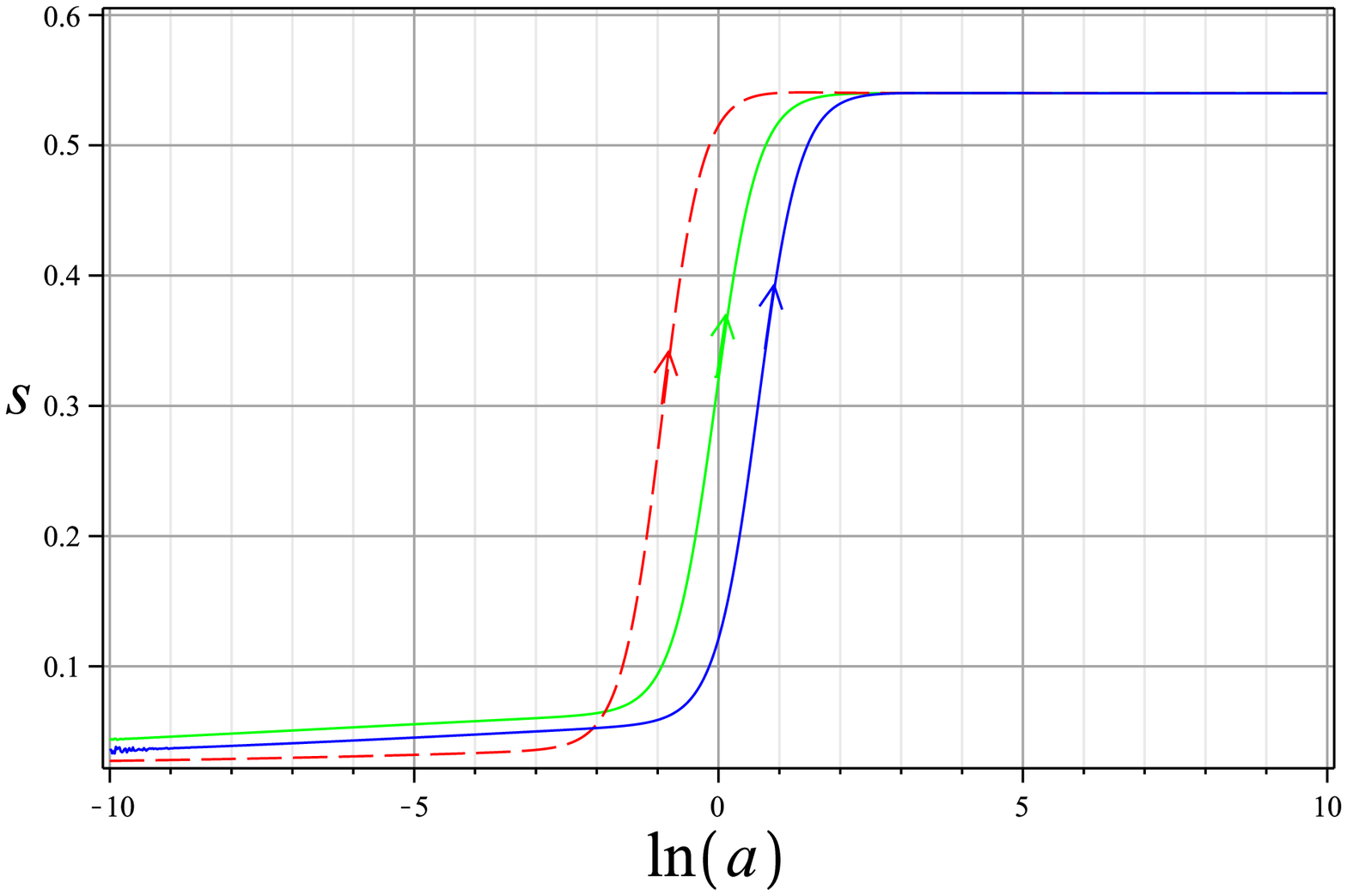}\hspace{0.1 cm}\includegraphics[scale=.4]{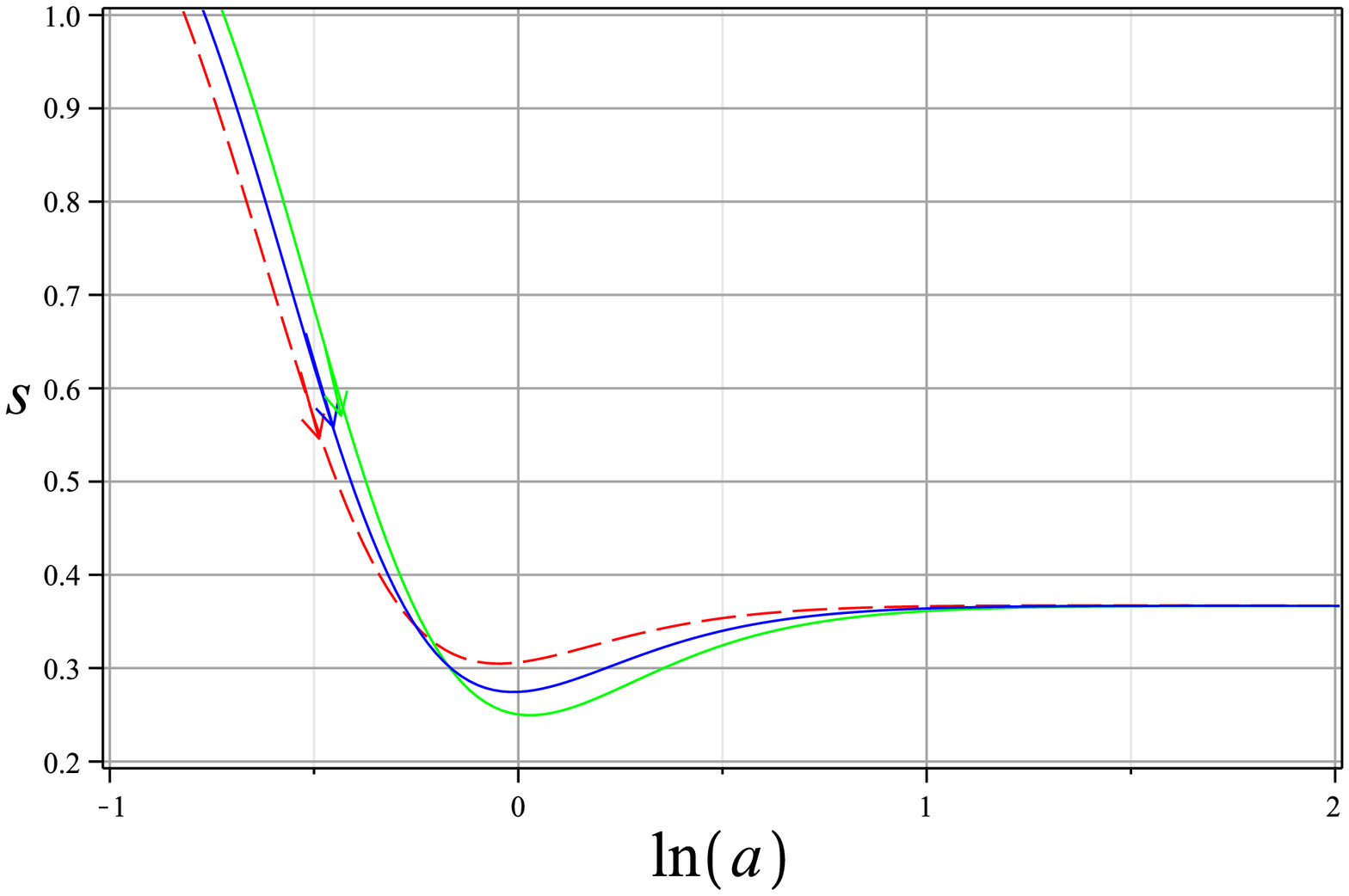}\hspace{0.1 cm}\\
Fig.13: The graph of the best-fitted statefinder  $ ({s,\ln (a))}$ for left) $\gamma=0$, right) $\gamma=\frac{1}{3}$.\\ The red trajectory is for the both best-fitted model parameters and best I.C.s
\end{tabular*}\\

\subsection{CRD test}

The Cosmological Redshift Drift is extracted from
\begin{eqnarray}\label{dotz}
\dot{z}=(1+z)H_0-H(z),
\end{eqnarray}
known as Mc Vittie equation. The equation immediately leads to velocity drift
\begin{eqnarray}\label{vdrift}
\dot{v}=cH_0-\frac{cH(z)}{1+z}.
\end{eqnarray}
with respect to the redshift that, by using $H(z)$ from numerical computation in our model, can be obtained against observational data. In our model, $H(z)$ is taken from numerical calculation in terms of the stability dynamical variables and after best fitting the model with the stability parameters and initial conditions. Fig. 14 shows a comparison between the velocity drift in our model for $\gamma =0, 1/3$ and $\Lambda$CDM model. As can be seen our model for $\gamma =0$ better fits the observational data in comparison with $\Lambda$CDM mode. It also shows another disadvantage in $\gamma =1/3$ case.\\

\begin{tabular*}{2.5 cm}{cc}
\includegraphics[scale=.5]{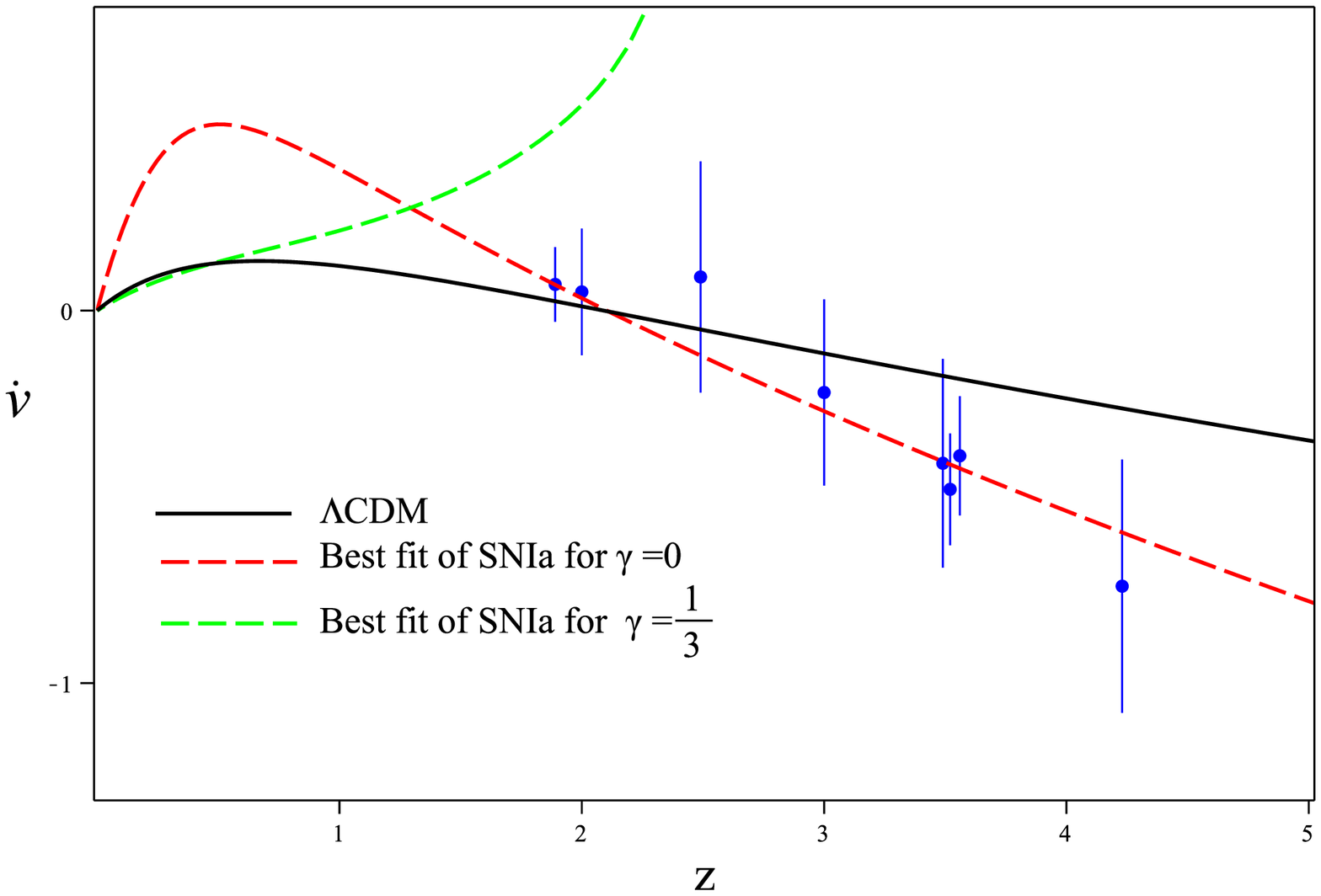}\\
Fig. 14: The graph of velocity drift as function of redshift for $\gamma=0$ and $\gamma=\frac{1}{3}$
\end{tabular*}\\

\section{Summary and reamrks}

This paper is designed to study the attractor solutions of a cosmological model with tachyonic potential and non minimally coupled scalar field with the matter, by stability analysis and making use of
 the 3-dimensional phase space of the theory. The model
 characterized by the scalar field $\phi$
 , the scalar potential $V(\phi)$, and the scalar function $f(\phi)$ nonminimally coupled to the matter lagrangian. The scalar function $f(\phi)$ and potential $V(\phi)$ are arbitrary and not specified in advance. The stability analysis gives the corresponding conditions
for tracking attractor and determines the type
of the universe behavior in the past and future.

In this work, with a new approach in stability analysis, we simultaneously best-fitting the stability parameters with the observational data using $\chi^2$ method and solving the system of equations. The advantage of this approach is that the obtained critical points are observationally verified and thus physically more promising. The best fitted critical points are physical critical points. By this approach in our model we find that one of the mathematically known stable critical point (FP1) is now physically unstable. Another advantage of this approach is that the fixed points coordinates are best fitted. We added even one more step in the process and best-fitted both stability parameters and initial conditions in the model. As a result, we obtained the observationally verified trajectories in the phase plane.

We then study the cosmological parameters such as effective EoS parameter, $\omega_{eff}$, deceleration parameter, $q$, and statefinder parameters
for the model in terms of the  best-fitted stability parameters. It shows that the EoS parameter for both $\gamma=0, 1/3$, does not cross the cosmological divide line in the past and future. In $\gamma=0$ case, the universe starts from unstable state in the past with $\omega_{eff}=0$ and finally tends to the stable state in the future with $\omega_{eff}=-0.47$. The current value of the EoS parameter of the universe for both best fitted stability parameters and initial conditions is within the range of observationally accepted values. The model also tested with CRD observational data and shows a better match with the data compare to the $\Lambda$CDM model and also case $\gamma=1/3$. With both best-fitted stability parameters and initial conditions, the deceleration parameter $q$ satisfies $q < 0$ at present and $q>0$ in the matter dominated era, in the case of $\gamma=0$. The statefinder parameters show that the universe start from SCDM in the past and approaches a state near SS in future for $\gamma=0$. In $\gamma=1/3$ case, the model is not observationally verified as the universe starts from an accelerated state in the past and the EoS parameter also begins from an unstable state with $\omega_{eff}=-\infty$. In addition, the model does not explain the CRD data. Note that in this case, from stability equations (\ref{x11}-\ref{y11}), the new dynamical variables are independent of one of the stability parameters, $\alpha$. Since $\alpha$ is proportional to ${\frac{\dot{f}}{f}}$, the scenario is similar to a cosmological model with no coupling between the scalar field and matter which can justify the unsatisfactory result in $\gamma=1/3$ case.

\acknowledgments{The authors would like to thank the anonymous referee for valuable comments
to significantly improve the manuscript. We would also like to thank the research council of University of Guilan for financial support.}

\end{document}